\documentclass{aastex}
\usepackage{spr-astr-addons}
\usepackage{graphicx}

\RequirePackage{color}
\def\note #1]{{\bf #1]}}
\def\fig{.}
\def\dd{{\rm d}}
\def\Msun{\,{\rm M}_\odot}
\def\Xs{X_{\rm s}}
\def\Zs{Z_{\rm s}}
\def\bolddelta{\delta\kern-0.45em\delta\kern-0.45em\delta}
\def\boldr{\mbox{\boldmath$r$}}
\def\bolddelr{\bolddelta \boldr}
\def\muHz{\,\mu{\rm Hz}}
\def\CE{{\cal E}}
\newcommand{\emaila}{jcd@phys.au.dk}

\begin{document}

\title{Prospects for asteroseismology}
\shorttitle{Prospects for asteroseismology}
\shortauthors{Christensen-Dalsgaard and Houdek}

\author{J{\o}rgen Christensen-Dalsgaard} 
\affil{Department of Physics and Astronomy, Aarhus University,\\
DK-8000 Aarhus C, Denmark}
\and
\author{G\"unter Houdek} 
\affil{Institute of Astronomy, University of Vienna, A-1180 Vienna, Austria}
\email{\emaila}


\begin{abstract}
The observational basis for asteroseismology is being
dramatically strengthened, through more than two years of data from the
CoRoT satellite, the flood of data coming from the Kepler mission and,
in the slightly longer term, from dedicated ground-based facilities.
Our ability to utilize these data depends on further development of
techniques for basic data analysis, as well as on an improved 
understanding of the relation between the observed frequencies and the
underlying properties of the stars.
Also, stellar modelling must be further developed, to match the
increasing diagnostic potential of the data.
Here we discuss some aspects of data interpretation and modelling,
focussing on the important case of stars with solar-like oscillations.
\end{abstract}



\section{Introduction}

\label{sec:intro}
%
In the last two decades data have become available on solar oscillations,
which have allowed detailed inferences to be made on the properties of the
solar interior \citep{Christ2002}.
This has led to the determination of the solar sound speed with considerable
accuracy, detailed tests of the equation of state of the solar interior and 
a determination of the solar internal rotation, the origin of which is still
poorly understood. The inferences about the solar internal structure obviously
provide a strong test of modelling of stellar evolution.
Although earlier solar models were in reasonable, but far from perfect,
agreement with the helioseismic determinations \citep{Gough1996},
recent redeterminations of the solar surface abundance have led to a very
substantial increase in the discrepancies between the Sun and the model
\citep[for reviews, see][see also Section~\ref{sec:soundspeed}]
{Guzik2006, Guzik2008, Basu2008}.

The Sun is only one star at a specific stage of its evolution, and its 
structure is relatively simple, compared with other stars.
Thus it is of obvious importance to extend seismic investigations to 
other stars.
This is becoming possible, thanks to the development of ground-based
instrumentation and the launch of three space missions that are providing
extensive data on stellar oscillations.
Thus it is important to consider the tools available for the analysis of
such asteroseismic data and summarize the results that have already been
obtained.



Here we concentrate on solar-like oscillations;
these are characterized by being intrinsically damped and excited stochastically
by near-surface convection
\citep[e.g.,][]{Houdek1999, Houdek2006}.
Thus they are found in relatively cool stars, with significant outer convection
zones, both on the main sequence and on the red-giant branch, as recently
confirmed spectacularly by the CoRoT mission
\citep{DeRidd2009}.
The modes relevant to observations of distant stars have low spherical-harmonic
degree and hence are generally high-order acoustic modes;
however, in evolved stars they may take on a mixed character, with a 
component corresponding to an internal gravity mode in the deep interior 
of the star, very substantially enhancing their diagnostic value
\citep[e.g.,][see also Section~\ref{sec:mixmodes}]{Christ2004}.


At the simplest level, the oscillation frequencies are determined by the 
global properties of the star, more specifically the mean density
$\bar \rho \propto M/R^3$, where $M$ is the mass and $R$ the surface radius
of the star.
When combined with additional, nonseismic, data and stellar modelling,
the stellar radius can be determined with substantial precision and,
one may hope, accuracy \citep{Stello2009}.
Such a determination of the radius is particularly important for 
observations of extra-solar planets using the transit technique,
where determination of the radius of the planet requires knowledge of
the stellar radius \citep{Kjelds2009}.
From the point of view of stellar modelling, however, it is of substantially
greater interest to obtain information about the stellar interior.
From low-degree acoustic modes a measure can be obtained that is sensitive
to the properties of the stellar core and hence, with additional assumptions,
provides a measure of the amount of hydrogen consumed during stellar evolution
and thus of the stellar age.
With sufficiently detailed and accurate observations additional information
can be obtained about the stellar interior, including inferences of the
sound-speed and density variation in the central parts of the star
\citep[e.g.,][]{Gough1993a, Basu2002, Roxbur2004}.

%
%
%

An obvious requirement for asteroseismic inferences is the availability of
good observational data.
This is increasingly being met (see Section~\ref{sec:obs}).
Ground-based observations of Doppler velocity have been possible for a few
stars, at level of sensitivity of a few ${\rm cm \, s^{-1}}$.
Space-based asteroseismic observations were started serendipitously 
by the WIRE satellite \citep{Buzasi2005, Bruntt2007},
followed by the launch in 2003 of the highly successful
Canadian micro-satellite MOST \citep{Walker2003, Matthe2007}.
The French-led CoRoT satellite mission \citep{Baglin2009},
launched at the end of 2006, has provided 
photometric oscillation data on a large number of stars, and high-quality data
on an even more extensive stellar sample is promised by the NASA Kepler mission
launched in March 2009 \citep{Boruck2009}.
Further observational projects promise a substantial increase in
the quality of data available for asteroseismic analysis. 
However, the interpretation also depends on an adequate understanding
of the relation between the observed oscillation properties, in particular
the frequencies, and the underlying properties of the star, in order to
design optimal diagnostics based on the observations.
Furthermore, the physical interpretation of the results relies on stellar
modelling, relating the physics of stellar interiors to stellar structure.
Thus a crucial aspect of the analysis is the reliable computation of stellar
models.
This should obviously ensure numerical accuracy;
an important step towards this goal is the detailed comparison of stellar
models and oscillation frequencies computed by independent codes
that has been undertaken in the ESTA collaboration,
initially as part of the CoRoT project
\citep{Lebret2008a, Lebret2008b, Moya2008}.
However, an equally important aspect is to ensure that the unavoidable
approximations in the modelling are implemented in a consistent
and accurate manner;
only in this case can the asteroseismic analysis provide a reliable
assessment of these approximations and, ideally, point to ways to improve them.
Finally, the results should be presented in a statistically well-defined
manner, such that the significance of the inferred properties, and
of any deviations from the assumed models, can be assessed.
It is probably fair to say that we are still some way from meeting these
requirements.

In the present paper we discuss some aspects of the diagnostics of
solar-like oscillations, based on asymptotic relations and
computed properties; this in particular emphasizes the important probing 
of stellar cores.
In addition, we consider some examples of uncertainties in stellar 
internal physics which have been, or may be, amenable to seismic inferences.
More extensive overviews of helio- and asteroseismic diagnostics were
provided by, for example, \citet{Gough1993b} and \citet{Christ2004},
and a comprehensive discussion of asteroseismology can be found
in \citet{Aerts2009}.

\section{Asteroseismic diagnostics}

\subsection{Simple asymptotics of acoustic modes}

\label{sec:simpleas}
%
The observed oscillation modes in main-sequence solar-like stars are 
driven predominantly by the vigorous turbulence in the superficial 
stellar layers. This results in a rich acoustic pulsation spectrum with 
a characteristic comb-like frequency structure. In distant stars only 
low-degree modes can be observed; however, the observed modes are generally 
of high order allowing us to use asymptotic theory. For modes
satisfying $l/n\rightarrow0$, where $n$ is radial order and $l$ degree,
the frequencies $\nu_{n l}$ can be estimated as
\citep{Vandakurov1967, Tassoul1980, Gough1986}:
\begin{equation}
\nu_{n l}\simeq\left(n+\frac{l}{2}+\epsilon\right)\nu_0
-\frac{AL^2-B}{\nu_{n l}}\nu^2_0+{\rm O}(\nu^4_0)\,,
\label{eq:asymptotics}
\end{equation}
where
\begin{equation}
\nu_0=\left[2\int_0^R\frac{{\rm d}r}{c}\right]^{-1}
\end{equation}
is the inverse of twice the sound travel time between the centre and 
acoustic surface, $c$ is the sound speed, and
\begin{equation}
A=\frac{1}{4\pi^2\nu_0}\left[\frac{c(R)}{R}
 -\int_0^R\frac{{\rm d}c}{{\rm d}r}\frac{{\rm d}r}{r}\right]\,.
\label{eq:biga}
\end{equation}
The frequency-dependent coefficient $\epsilon$ is determined by the 
reflection properties of the surface layers, as is the small correction term
$B$ \citep[e.g.,][]{Gough1986}, and $L=\sqrt{l(l+1)}$. The value of $\nu_0$ can 
be estimated from taking the average (over $n$ and $l$) of the so-called large 
frequency separation
\begin{equation}
\Delta\nu_{n l}\equiv\nu_{n+1\, l}-\nu_{n l}
\label{eq:largesep}
\end{equation}
between modes of like degree and consecutive order. The resulting comb-like
frequency structure consists of modes of odd degree falling approximately 
halfway between modes of even degree.
The last two terms on the right-hand side of equation (\ref{eq:asymptotics}) 
lift the degeneracy between modes with the same value of $n+l/2$, leading
to the so-called small frequency separation 
$\delta\nu_{n l}\equiv\nu_{n l}-\nu_{n-1\,l+2}$. The small frequency separation
measures the extent to which the integrated history of nuclear transmutations 
has modified the structure of the energy-generating core.
The mean small frequency separation, averaged over $n$ (indicated by 
angular brackets) becomes in the limit $l/n\rightarrow0$
\begin{equation}
\langle\delta\nu_{n l}\rangle
=\langle\nu_{n l}-\nu_{n-1\,l+2}\rangle\,
\simeq\,2A(2l+3)\frac{\nu^2_0}{\langle\nu_{n l}\rangle}\,,
\label{eq:meansmallsep}
\end{equation}
and is therefore proportional to the coefficient $A$ (Eq.~\ref{eq:biga}).
The coefficient $A$ measures predominantly the stratification of the 
energy-generating core through the gradient of the sound speed and
hence is sensitive to the chemical composition there and
consequently is an indicator for stellar age 
\citep[e.g.,][]{Christ1984, Christ1993, Gough1986, Gough2001, 
Ulrich1986, Houdek2008}.

From Eq.~(\ref{eq:asymptotics}) it follows that, 
to this asymptotic order,
\begin{equation}
D_{nl}= {\nu_{nl} - \nu_{n-1\,l+2} \over 2 l + 3} 
\label{eq:scalesep}
\end{equation}
is a function of frequency.
The same is true for the following ratio:
\begin{equation}
D_{nl}^{(1)} = 
{(\nu_{n-1\,l} + \nu_{nl})/2 - \nu_{n-1 \, l+1} \over l + 1} \; ,
\label{eq:scalesep1}
\end{equation}
which is also a useful diagnostic, particularly when only modes
of degree 0 and 1 can be identified from observations.

\subsection{The inner phase shift}

\begin{figure}[t]
\centering
\includegraphics[width=0.5\textwidth]{\fig/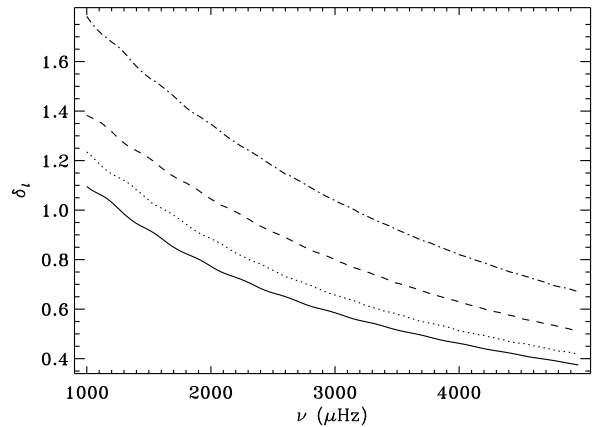}
\caption{%
The internal phase shift $\delta_l$ (cf.\ Eq.~\ref{eq:asfunc})
for a model of the present Sun, as a function of cyclic frequency.
Continuous curve: $l = 0$;
dotted curve: $l = 1$;
dashed curve: $l = 2$; and
dot-dashed curve: $l = 3$.
\label{fig:sunphase}
} 
\end{figure}

%
\label{sec:inphase}
Further insight into the diagnostic potential of low-degree solar-like
oscillations has been developed by Roxburgh and Vorontsov.
They represent the effect of the structure of the core by 
an internal phaseshift $\delta_l(\omega)$ \citep{Roxbur1994},
defined such that the asymptotic behaviour of the acoustic modes
can be expressed as
\begin{equation}
\frac{(\rho c)^{1/2}}{r} p'
\simeq A_{\rm p} \sin\left(\omega \tilde\tau - \frac{\pi}{2} l + \delta_l\right) \; ,
\label{eq:asfunc}
\end{equation}
where $\rho$ is density, $p'$ is the Eulerian perturbation to pressure,
$\omega = 2 \pi \nu$ is angular frequency and
\begin{equation}
\tilde \tau = \int_0^r {\dd r \over c}
\label{eq:acoustic-rad}
\end{equation}
is the acoustic distance from the centre, corresponding to the radius $r$.
They noted that $\delta_l$ can be computed as a function of frequency
by matching partial solutions to the oscillation equations to
the asymptotic expression in Eq.~(\ref{eq:asfunc}).
We have implemented this by considering solutions satisfying both the
central and surface boundary conditions, but allowing a discontinuity 
in the solution at a suitably chosen point near the surface.
Figure~\ref{fig:sunphase} shows the results for a model of the present Sun.

From Eq.~(\ref{eq:asfunc}) and a similar asymptotic representation of
the solution in the outer parts of the model, 
the oscillation frequencies can be obtained as
\begin{equation}
\nu_{nl} =  \nu_0 \left[ n + {l \over 2}
+ \pi^{-1} (\tilde\alpha(\omega_{nl}) - \delta_l(\omega_{nl})) \right] \; ,
\label{eq:rvasfreq}
\end{equation}
where $\tilde\alpha(\omega)$ is a surface phase function
\citep[see also][]{Christ1992} which, for low-degree modes, is
independent of degree.
This expression is clearly closely related to the simple asymptotic 
expression in Eq.~(\ref{eq:asymptotics}) above,
with the small frequency separation being related to $\delta_l$ through
\begin{equation}
\nu_{nl} - \nu_{n-1\, l+2} = {\nu_0 \over \pi} (\delta_{l+2} - \delta_l)
\; .
\label{eq:rvsmallsep}
\end{equation}

\begin{figure}[t]
\centering
\includegraphics[width=0.5\textwidth]{\fig/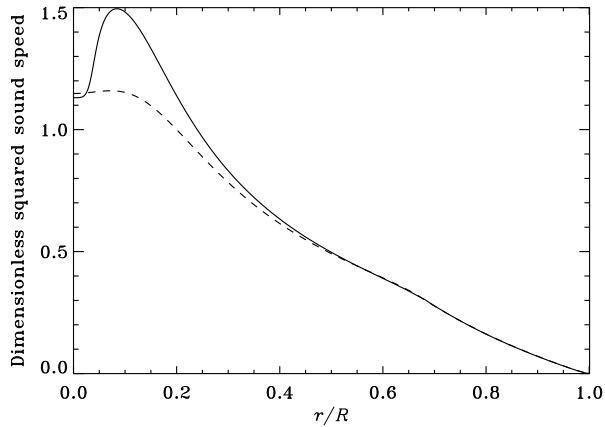}
\caption{%
Dimensionless squared sound speed, in units of $GM/R$,
as a function of fractional radius,
for models of $\alpha$~Cen~A (solid) and B (dashed)
\label{fig:csqacen}
} 
\end{figure}


\citet{Roxbur1994} noted that, in the simplest asymptotic approximation,
$\delta_l$ can be expressed as
\begin{equation}
\delta_l(\omega) = \delta_0(\omega) + l(l+1) \hat \delta(\omega) \; ,
\label{eq:rvphase}
\end{equation}
where
\begin{equation}
\hat \delta \simeq {1 \over 2 \omega}
\left[ {c(R) \over R} - \int_0^R {\dd c \over \dd r} {\dd r \over r} \right]
\; .
\label{eq:rvdell}
\end{equation}
Given Eq.~(\ref{eq:rvsmallsep}), this evidently leads to
Eq.~(\ref{eq:meansmallsep}) for the average small separation, with $A$
given by Eq.~(\ref{eq:biga}).

\begin{figure}[t]
\centering
\includegraphics[width=0.5\textwidth]{\fig/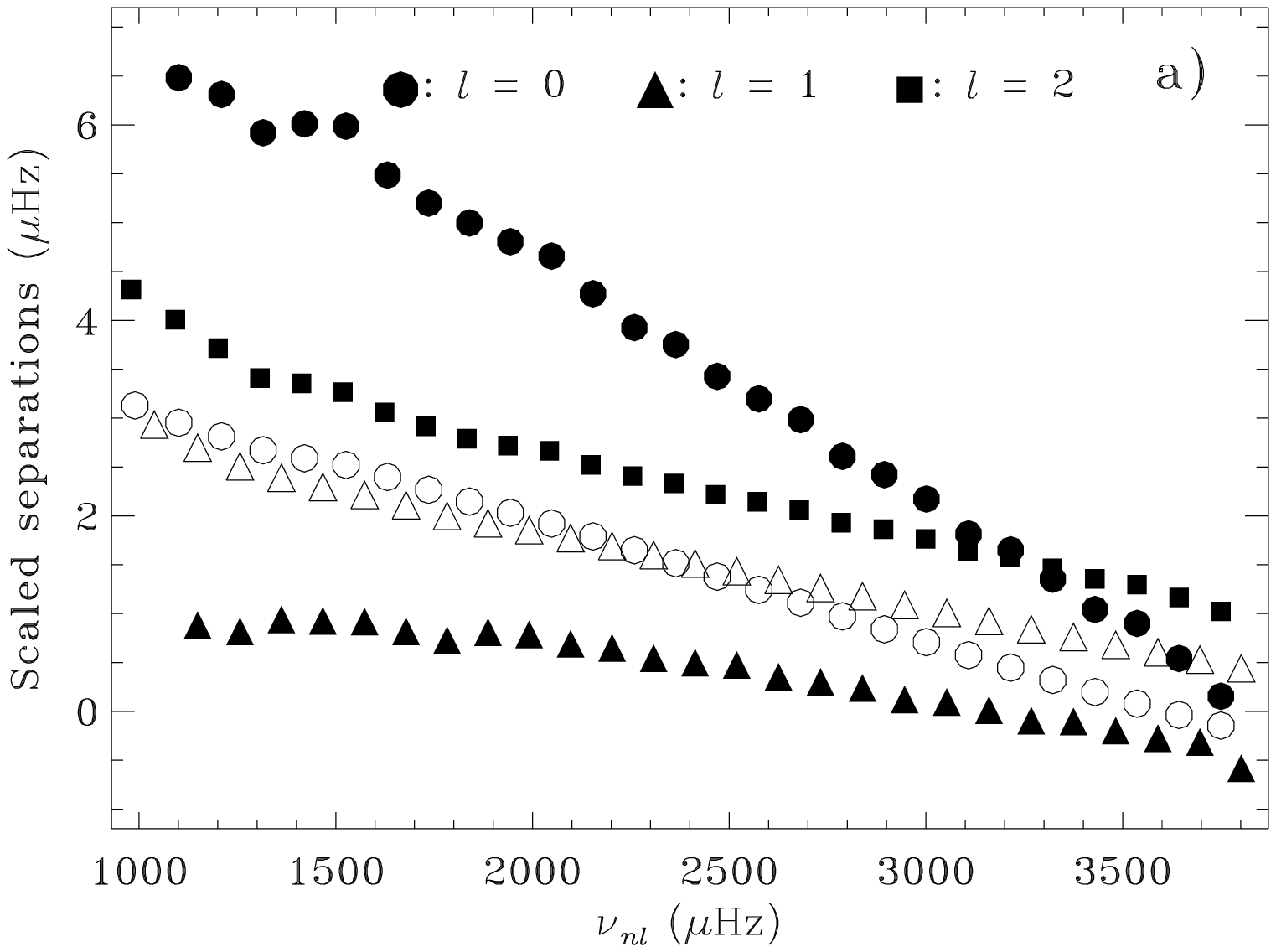}
\includegraphics[width=0.5\textwidth]{\fig/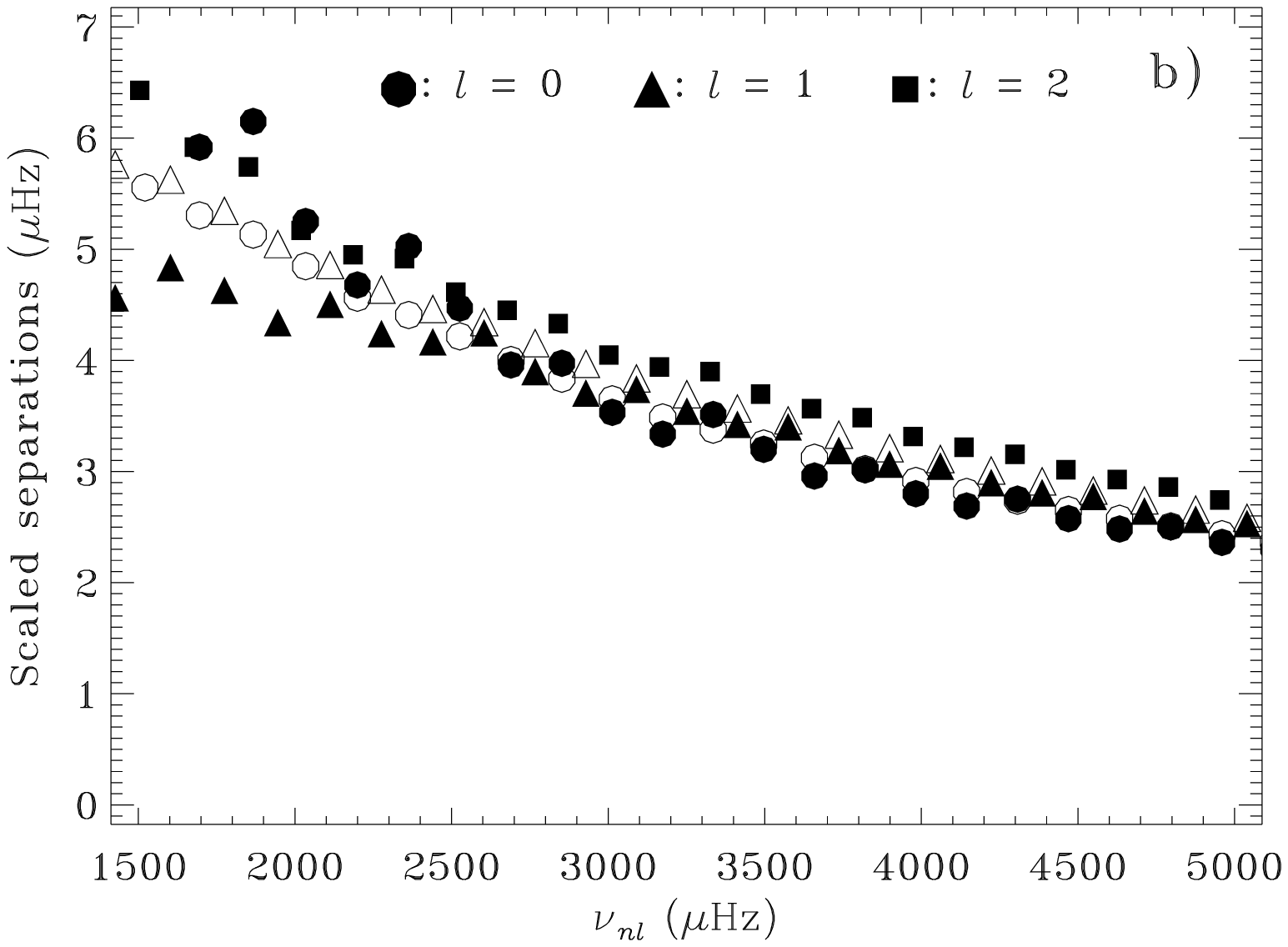}
\caption{%
Scaled small frequency separations (cf.\ Eqs \ref{eq:scalesep} 
and \ref{eq:scalesep1}),
for models of $\alpha$ Cen A (panel a) and B (panel b).
Filled symbols:
$D_{nl}^{(1)}$, for $l = 0$, 1 and 2; the degree is indicated by
the symbol, as shown in the figure.
Open symbols: 
$D_{nl}$, for $l = 0$ and 1.
Note that according to Eq.~(\ref{eq:asymptotics})
(see also Eq.~\ref{eq:meansmallsep}) these quantities are expected to
be functions of frequency alone
\label{fig:scsepacen}
} 
\end{figure}

\begin{figure}[t]
\centering
\includegraphics[width=0.5\textwidth]{\fig/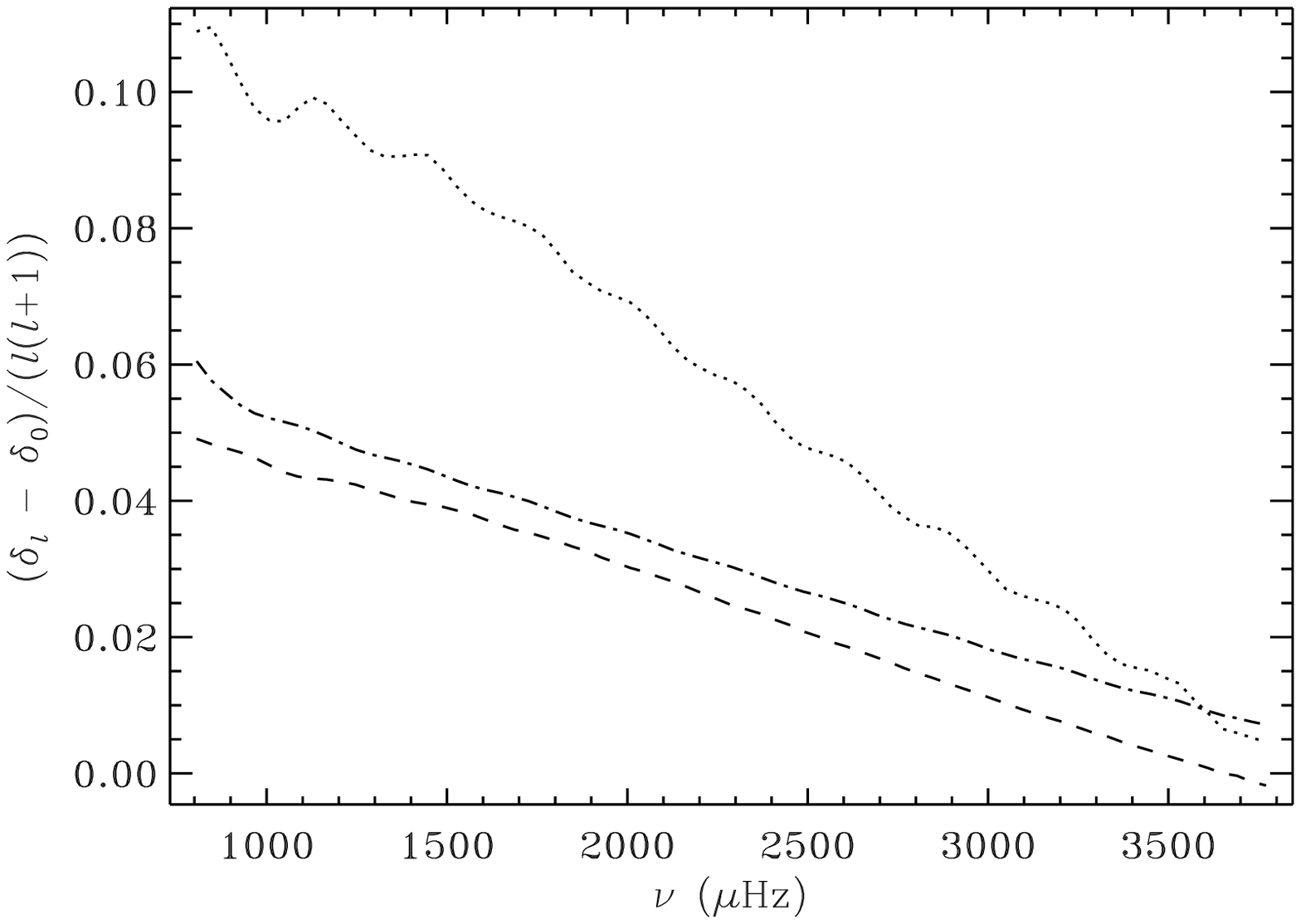}
\includegraphics[width=0.5\textwidth]{\fig/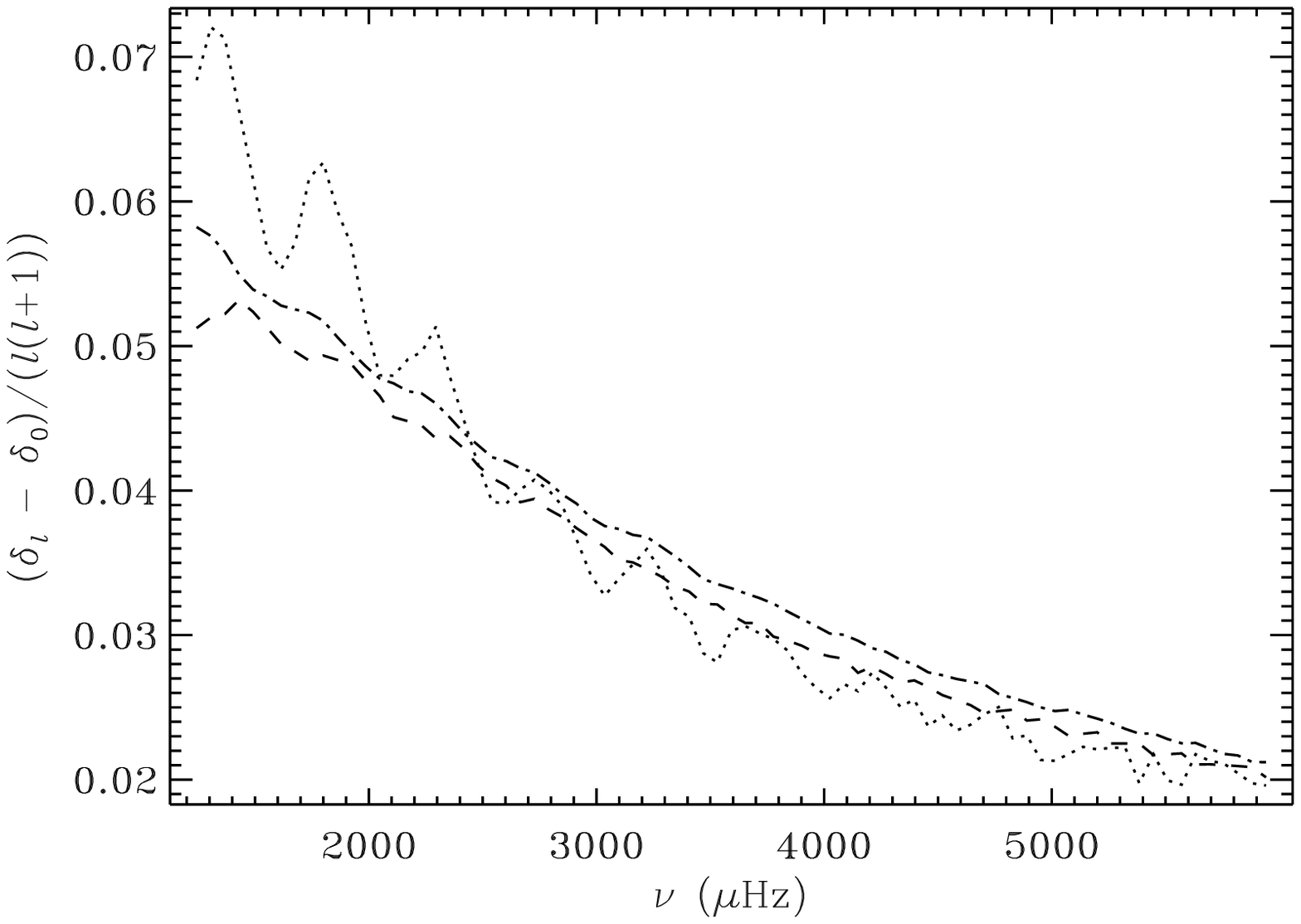}
\caption{%
Scaled differences in internal phase shift,
for models of $\alpha$ Cen A (top) and B (bottom).
Dotted curve: $l = 1$;
dashed curve: $l = 2$;
dot-dashed curve: $l = 3$.
Note that according to Eq.~(\ref{eq:rvphase}) these differences are expected
to be functions of frequency alone
\label{fig:scdphase}
} 
\end{figure}

The simple asymptotic behaviour in Eqs~(\ref{eq:asymptotics}) and
(\ref{eq:biga}), or Eq.~(\ref{eq:rvphase}) and Eq.~(\ref{eq:rvdell}),
assumes that the equilibrium structure is varying slowly with position.
As a star evolves on the main sequence, the increase in the central
mean molecular weight leads to a local decrease in the sound speed,
as illustrated in Fig.~\ref{fig:csqacen} 
for the components of $\alpha$ Cen binary system.
At the age of 6.98\,Gyr of the system the B component, with a mass of
$0.928 \Msun$, is relatively unevolved, while the A component, with
a mass of $1.111 \Msun$, is nearing the end of central hydrogen burning.
The effect is evident in the difference in the behaviour of
the dimensionless sound speed
and is reflected in the scaled small separations
shown in Fig.~\ref{fig:scsepacen}.
While the scaled separations, in accordance with Eq.~(\ref{eq:asymptotics}),
are essentially just a function of frequency for $\alpha$ Cen B,
for the A component there is very substantial scatter, reflecting
the rapid variation of the sound speed in the core.

This behaviour is illustrated in Fig.~\ref{fig:scdphase},
in terms of the scaled
differences $[l(l+1)]^{-1}(\delta_l - \delta_0)$ in the internal phase shift,
which according to Eq. (\ref{eq:rvphase}) is expected to be a function
of frequency.
This is clearly approximately satisfied for the B component, whereas
for the A component the results for $l = 1$ deviate strongly from
the expected asymptotic behaviour.
Interestingly, the differences for $l = 1$ show a distinct 
oscillatory behaviour which is related to 
the base of the convection zone 
\citep[see also][and Section~\ref{sec:glitch}]{Roxbur2009}.

\begin{figure}[t]
\centering
\includegraphics[width=0.5\textwidth]{\fig/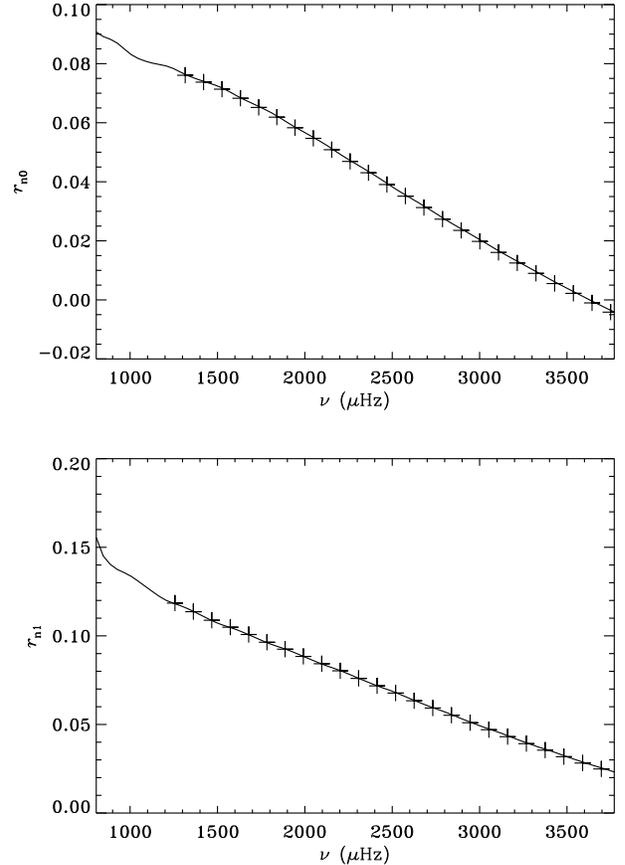}
\caption{%
Frequency-separation ratios (cf.\ Eq.~\ref{eq:seprat}; symbols) and
difference $(\delta_{l+2} - \delta_l)/\pi$ in internal phase shifts (curve),
for a model of $\alpha$ Cen A.
Top: $l = 0$, bottom: $l = 1$.
\label{fig:seprat}
} 
\end{figure}

It was noted by \citet{Roxbur2003} that the uncertainty in the
modelling of the near-surface layers (see Section~\ref{sec:surfprob})
has a significant effect on the small frequency separation.
They pointed out that this effect is strongly suppressed by
considering instead ratios such as
\begin{eqnarray}
&& r_{n0} = {\nu_{n0} - \nu_{n-1\,2} \over \nu_{n-1\,1} - \nu_{n1}} 
\nonumber \\
&& r_{n1} = {\nu_{n1} - \nu_{n-1\,3} \over \nu_{n0} - \nu_{n+1\,0}} \; .
\label{eq:seprat}
\end{eqnarray}
They also demonstrated that to this accuracy
the ratios are related to the internal phase shifts by
\begin{equation}
r_{n0} = (\delta_2 - \delta_0)/\pi \; , \quad
r_{n1} = (\delta_3 - \delta_1)/\pi \; .
\end{equation}
This is illustrated in Fig.~\ref{fig:seprat} for a model of $\alpha$ Cen A.
The diagnostic potential of these separation ratios, and their
insensitivity to even very substantial differences in the surface layers,
was further analyzed by \citet{Oti2005} and \citet{Roxbur2005}.

\begin{figure}[t]
\centering
\includegraphics[width=0.5\textwidth]{\fig/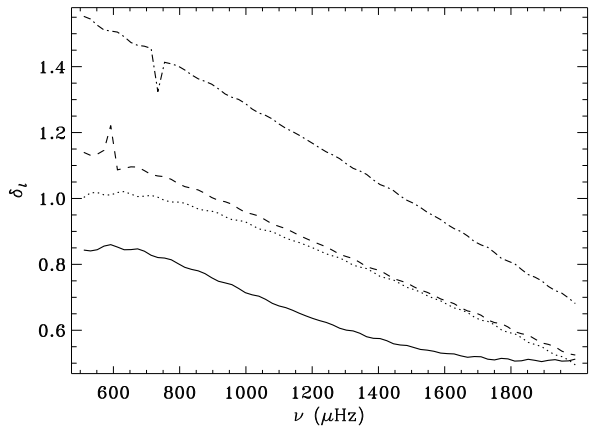}
\caption{%
The internal phase shift $\delta_l$ for a model approximating Procyon, 
with mass $M = 1.5 \Msun$ and central hydrogen abundance $X_{\rm c} = 0.05$.
Continuous curve: $l = 0$;
dotted curve: $l = 1$;
dashed curve: $l = 2$; and
dot-dashed curve: $l = 3$.
The glitches for $l = 2$ and 3 are probably associated with mixed modes
(see Section~\ref{sec:mixmodes})
\label{fig:procphase}
} 
\end{figure}

One may hope that further analysis of the internal phase function can provide
insight on the diagnostic potential of the frequency separations, beyond the
simplest asymptotics.
\citet{Roxbur1994} developed relatively simple expressions for $\delta_l$,
in terms of the structure of the core, although these have apparently not
been exploited in detail so far \citep[see also][]{Roxbur2000, Roxbur2007}.


As discussed in Section~\ref{sec:concore} the diagnostics of convective cores
is particularly important. 
The mixing in the core leads to sharp variations in composition and
hence sound speed  there, causing
strong departures from the simple asymptotics of low-degree modes, as 
reflected in the inner phase shift and the small frequency separations.
This is illustrated in Fig.~\ref{fig:procphase} for a model
resembling Procyon. 
A comparison with the results for the much more benign solar model,
illustrated in Fig.~\ref{fig:sunphase},
clearly shows the change in the behaviour of the inner phase shift $\delta_l$.
This indicates the potential for using frequencies of low-degree modes
to probe the properties of convective cores in stars showing solar-like
oscillations \citep[see also][]{Popiel2005}.
\citet{Cunha2007} considered the specific case of a small growing convective
core, deriving diagnostics, based on an asymptotic analysis of
the oscillations, of the properties of the core. 
They found that the quantity 
\begin{equation}
\Delta_{\rm CM} = {D_{n0} \over \Delta \nu_{n-1 \, 1}}
- {D_{n1} \over \Delta \nu_{n \, 0}} 
\end{equation}
(see Eqs~\ref{eq:largesep} and \ref{eq:scalesep})
provides a measure of stellar age; more specifically, it is sensitive
to the discontinuity in the sound speed resulting from the composition
discontinuity at the edge of the core.

\subsection{Diagnostics of acoustic glitches}

\label{sec:glitch}
The asymptotic expression (\ref{eq:asymptotics}) is valid only if the spatial
scale of variation of the equilibrium state is everywhere much greater than
the inverse vertical wavenumber of the mode. But that condition is not
satisfied in the Sun and main-sequence solar-like stars: there is small-scale
variation associated with ionization of abundant elements and the near
discontinuity in low derivatives of the density at the base of the convection
zone. These abrupt variations induce small-amplitude oscillatory components
(with respect to frequency) in the spacing of the 
cyclic eigenfrequencies $\nu_{n l}$ of seismic oscillation and
consequently also in $\Delta\nu_{n l}$ and $\delta\nu_{n l}$.
We call such abrupt variations an acoustic glitch.

A convenient and easily evaluated measure of the oscillatory component 
produced by acoustic glitches is the second multiplet-frequency difference 
with respect to order $n$ amongst modes of like degree $l$:
\begin{equation}
\Delta_2\nu_{n l}\equiv\nu_{n-1\,l}-2\nu_{n\,l}+\nu_{n+1\,l}
\label{eq:secdiff}
\end{equation}
\citep{Gough1990}. Any 
localized region of rapid variation of either the sound speed $c$ or the 
density scale height, or a spatial derivative of them, induces an 
oscillatory component in $\Delta_2\nu$ (the subscripts $n l$ 
have been dropped) with a 
`cyclic frequency' approximately equal to twice the acoustic depth
\begin{equation}
\tau=\int_{r_{\rm glitch}}^R c^{-1}\,{\rm d}r
\end{equation}
(not to be confused with the acoustic radius, Eq.\,\ref{eq:acoustic-rad}) 
of the glitch, 
and with an amplitude which depends on the amplitude of the 
glitch and which decays with $\nu$ once the inverse radial wavenumber of the 
mode becomes comparable with or less than the radial extent of the glitch.  

\begin{table}[t]
\begin{center}
\small
\caption{Seismically determined properties of acoustic glitches in the Sun
\citep[using BiSON data; ][]{Basu2007} and in Model S \citep{Christ1996} 
based on the analysis by \citet{Houdek2007} of low-degree modes $(l=0,...,3)$: 
$\delta\gamma_1/\gamma_1|_{\rm II}$ and  $\tau_{\rm II}$ are the relative 
depression of the first adiabatic exponent and the acoustic depth
of the second helium ionization zone, and $\tau_{\rm c}$ is the 
acoustic depth of the sharp transition from radiative to convective 
heat transport at the base of the convection zone
\label{tab:1}}
\begin{tabular}{@{}lccc@{}}
\tableline
      &$-\delta\gamma_1/\gamma_1|_{\rm II}$&$\tau_{\rm II}$ (s)&$\tau_{\rm c}$ (s)\\
\tableline
Sun    &0.043                          &819                &2310              \\
Model S&0.045                          &815                &2270              \\
\tableline
\end{tabular}
\end{center}
\end{table}

Various approximate formulae for the oscillatory components 
that are associated with the helium ionization have been suggested and used, 
by e.g., \citet{Basu1994, Basu2004}, \citet{Monteiro1998, Monteiro2005}, 
and \citet{Gough2002}, not all of which
are derived directly from explicit acoustic glitches. Gough used an
analytic function for modelling the dip in the first adiabatic exponent
$\gamma_1=(\partial\ln p/\partial\ln\rho)_s$, where $p, \rho$ and $s$ are 
pressure, density and specific entropy.
In contrast, Monteiro \& Thompson assumed a triangular form.
Basu et al. have adopted a seismic signature for helium ionization that is 
similar to that arising from a single discontinuity;
the artificial discontinuities in the sound speed and its
derivatives that this and the triangular representations possess cause 
the amplitude of the oscillatory signal to decay with frequency too gradually,
although that deficiency may not be immediately noticeable within the limited 
frequency range in which adequate asteroseismic data are or will 
imminently be available.
More recently \citet{Houdek2007} proposed a seismic diagnostic in which
the variation of the first adiabatic exponent $\gamma_1$ in the helium 
ionization zone is represented with a pair of Gaussian functions. 
This correctly results in a decay of the amplitude of the seismic 
signature with oscillation frequency that is faster than that which 
the triangular and the single-discontinuity approximations imply, and 
also takes some account of
the two ionization states of helium. Moreover, \citet{Houdek2007} incorporated 
the acoustic cutoff frequency into the variation of the eigenfunction phase
with acoustic depth $\tau$, thereby improving the discrepancy between the
seismically inferred depths of the acoustic glitches and that of 
a corresponding stellar model. 

\begin{figure}[t]
\centering
\includegraphics[width=0.45\textwidth]{\fig/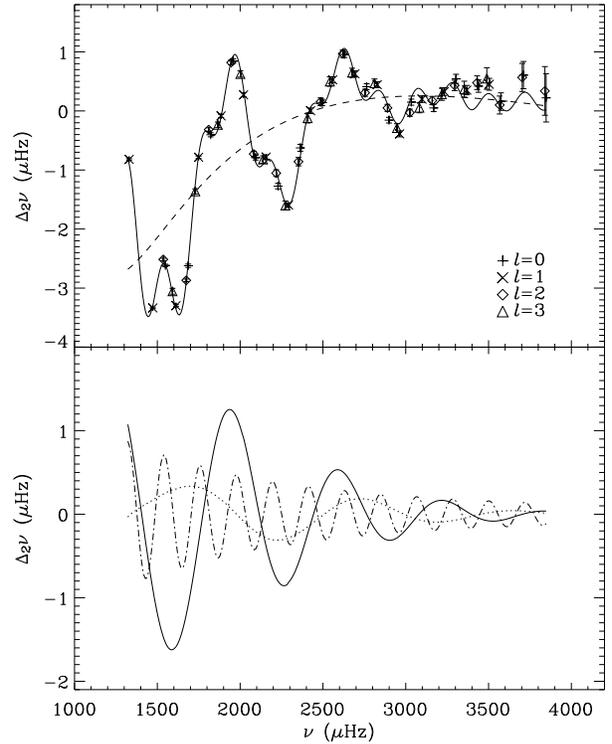}
\caption{%
{\bf Top}: Symbols show second differences $\Delta_2\nu_{n l}$ 
(Eq.\,\ref{eq:secdiff})
for low-degree modes of BiSON data  collected over a period of 
4752 days \citep{Basu2007}. The solid curve is Houdek \& Gough's (2007) 
seismic diagnostic, determined from fitting it to the data by least squares. 
The dashed curve is a smooth contribution of $\Delta_2\nu_{n l}$, modelled 
as a third-order polynomial in $\nu^{-1}$ guided by the asymptotic expression 
(\ref{eq:asymptotics}), and represents near-surface effects. 
{\bf Bottom:} Individual contributions of the seismic diagnostic. The solid 
curve
displays the contribution of the second ionization stage of the helium, and the 
dotted curve the first ionization stage of helium. The dot-dashed curve is the 
contribution from the discontinuity in the second density derivative at the 
base of the convection zone
} 
\label{fig:secdiff_sun}
\end{figure}

Fig.~\ref{fig:secdiff_sun} shows the 
outcome of fitting Houdek \& Gough's (2007) seismic diagnostic to 
BiSON data (top panel), together with the individual contributions from 
the acoustic glitches (lower panel).
The individual contributions provide a measure of the properties of the
acoustic glitches in the background stratification of the star.
One might hope that the variation of the sound speed $c$ induced by helium 
ionization might enable one to determine from the low-degree 
eigenfrequencies a measure that is directly related to, perhaps even almost 
proportional to, the helium abundance, with little contamination from other 
properties of the structure of the star.
Table~{\ref{tab:1}} compares some of the
properties of the acoustic glitches obtained from the BiSON data, 
and from adiabatically computed eigenfrequencies of Model~S 
\citep{Christ1996}, using the analysis by \citet{Houdek2007}.
The values between the Sun and Model~S differ by less than 3\%, 
demonstrating the diagnostic potential of this seismic analysis using only
low-degree modes.

\begin{figure}[t]
\centering
\includegraphics[width=0.45\textwidth]{\fig/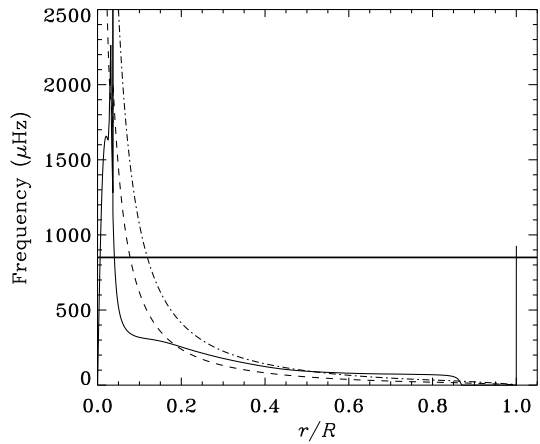}
\caption{%
Characteristic frequencies for a model of $\eta$ Boo.
The solid curve shows $N/2 \pi$ where $N$ is the buoyancy frequency
(cf.\ Eq.~\ref{eq:buoy}), and the dashed and dot-dashed curves show
$S_l/2 \pi$ for $l = 1$ and 2, respectively.
The heavy horizontal line marks a typical observed frequency of
$\eta$ Boo \citep[see][]{DiMaur2003}
\label{fig:eboochar}
} 
\end{figure}

\subsection{Mixed modes}

%
\label{sec:mixmodes}
We have so far considered only modes that are predominantly of acoustic nature;
these {\it p modes} are expected at the relatively high frequencies where the 
stochastic excitation is most efficient.
However, in evolved stars, beyond the end of central hydrogen burning,
internal gravity waves, or {\it g modes}, reach frequencies in the range
of the stochastically excited modes.
Their properties are characterized by the buoyancy, or Brunt-V\"ais\"al\"a,
frequency $N$, determined by
\begin{equation}
N^2 = g \left( {1 \over \gamma_1} {\dd \ln P \over \dd r} 
- {\dd \ln \rho \over \dd r} \right)
\simeq {g^2 \rho \over p} (\nabla_{\rm ad} - \nabla + \nabla_\mu) \; ,
\label{eq:buoy}
\end{equation}
where $g$ is the gravitational acceleration.
The last equality assumes the ideal gas law;
$\nabla = \dd \ln T/\dd \ln p$, $\nabla_{\rm ad}$ is its adiabatic value,
and $\nabla_\mu = \dd \ln \mu / \dd \ln p$, where $\mu$
is the mean molecular weight.
In evolved stars $g$ becomes very high near the compact core;
an additional contribution comes from $\nabla_\mu$ which is large in
the region of varying hydrogen abundance left behind after nuclear fusion. 
As a result, $N$ may reach very high values in the central parts of the star.
\begin{figure}[t]
\centering
\includegraphics[width=0.45\textwidth]{\fig/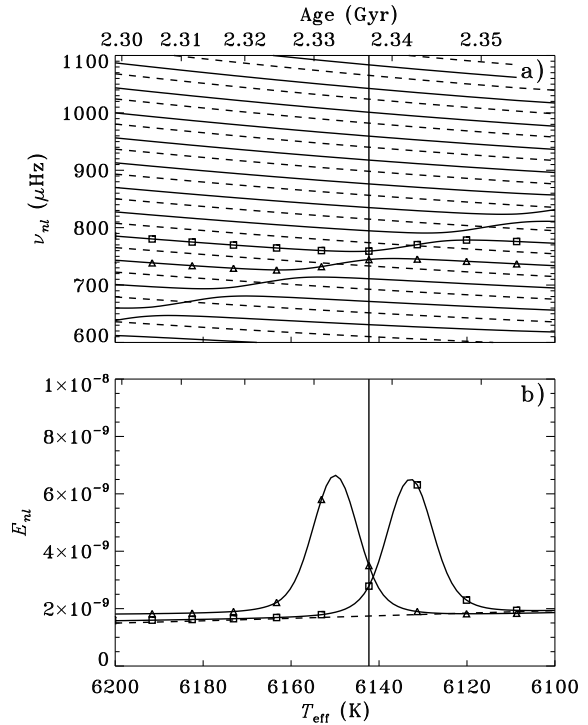}
\caption{%
Evolution of oscillation properties with age (upper abscissa) and
effective temperature (lower abscissa) in a model of $\eta$ Boo.
Panel (a) shows frequencies of modes with $l = 0$ (dashed curves)
and $l = 1$ (solid curves).
In panel (b) the solid curves show the evolution 
of the dimensionless inertia $E$
(cf.\ Eq.~\ref{eq:inertia}) for the two $l = 1$ modes marked
by triangles and squares in panel (a), and of a neighbouring mode with $l = 0$
(dashed curve).
The heavy vertical line identifies the model providing the best fit
to the observed frequencies.
\label{fig:eboofreq}
} 
\end{figure}

\begin{figure}[t]
\centering
\includegraphics[width=0.45\textwidth]{\fig/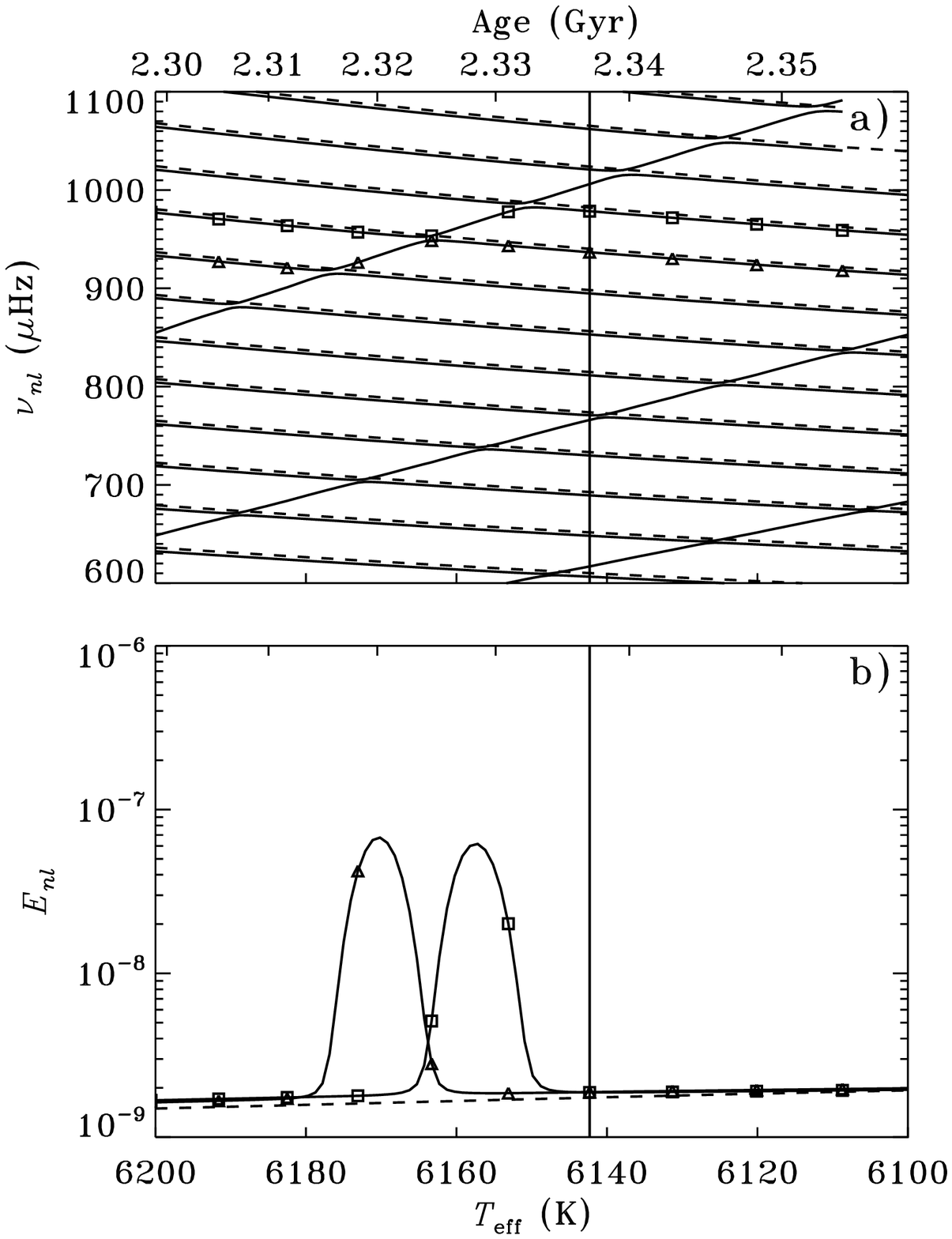}
\caption{%
As Fig.~\ref{fig:eboofreq}, but for $l = 2$.
\label{fig:eboofreq2}
} 
\end{figure}

As an example we consider a model of the subgiant $\eta$ Boo,
where for the first time individual frequencies of solar-like oscillations
in another star were identified \citep{Kjelds1995}.
This is based on modelling of the star by \citet{DiMaur2003};
the model has a mass of $1.7 \Msun$ and a heavy-element abundance $Z = 0.04$.
Figure~\ref{fig:eboochar} shows the buoyancy frequency at a model age
of 2.34\,Gyr.
It is evident that it reaches frequencies far higher than the 
acoustic cut-off frequency, around $1150 \muHz$, in the atmosphere of the model
which defines the upper limit to the frequencies of trapped acoustic modes.
Also shown are the acoustic, or Lamb, frequencies $S_1$ and $S_2$,
defined by 
\begin{equation}
S_l^2 = {l(l+1) c^2 \over r^2} \; .
\label{eq:lamb}
\end{equation}
The properties of a mode of oscillation are controlled by the magnitude of
the frequency relative to $S_l$ and $N$.
In the region of the star where the frequency exceeds $N$ and $S_l$
the mode behaves predominantly as an acoustic wave.
Where the frequency is below $N$ and $S_l$ the mode has the character 
of an internal gravity wave.
Finally, in the so-called evanescent region,
where the frequency is between $N$ and $S_l$, the mode amplitude varies
exponentially with position, either increasing or decreasing.

In Fig.~\ref{fig:eboochar} the horizontal line marks a typical
observed frequency in $\eta$ Boo.
It is evident that this mode may have g-mode behaviour in the core of
the model and p-mode behaviour in the envelope.
The dominant behaviour depends on the variation in the evanescent region.
If the eigenfunction decreases exponentially with $r$ in this region the mode
has predominantly the character of a g mode,
whereas if it increases exponentially, the mode has the character
of a p mode.
The discrimination between the two cases depends on the strength of the
exponential variation and hence on the extent of the evanescent region.
Thus the separation between the two cases is stronger for $l = 2$ 
than for $l = 1$ modes.

With the evolution on the sub-giant branch the star expands,
and hence the acoustic frequencies decrease approximately as $\bar \rho^{1/2}$.
On the other hand, the increasing central condensation increases the
buoyancy frequency and hence the frequencies of the g modes.
The combined effect for modes of degree $l = 0$ and 1, in models
near the identified stage of evolution of $\eta$ Boo, is illustrated
in Fig.~\ref{fig:eboofreq}.
For the radial modes, with $l = 0$, and the predominantly acoustic $l = 1$
modes the frequencies decrease with increasing age and hence decreasing
effective temperature.
However, there is evidently also a branch of frequencies, corresponding to
a mode of predominantly g-mode character, increasing with age.
Where this meets a predominantly acoustic mode, the two modes
undergo an {\it avoided crossing} \citep{Osaki1975},
where the frequencies approach quite closely without actually crossing. 
An example occurs at the vertical line, marking the model identified
as best fitting the observations.
At closest approach, the modes have a fully mixed nature, 
with similar amplitudes (in terms of energy density)
in the central g-mode region and the p-mode envelope.

The variation in the physical nature of the modes can be followed by
considering the normalized mode inertia
\begin{equation}
E = {\int_V \rho |\bolddelr|^2 \dd V \over M |\bolddelr_{\rm phot}|^2 } \; ,
\label{eq:inertia}
\end{equation}
where $\bolddelr$ is the displacement vector, $\bolddelr_{\rm phot}$
is its photospheric value, and integration is over the volume $V$ of the star.
This is illustrated in panel (b) of Fig.~\ref{fig:eboofreq} for two modes
with $l = 1$ undergoing an avoided crossing, and for a neighbouring
radial mode.
For the latter, a purely acoustic mode, $E$ varies slowly with age.
The $l = 1$ modes have $E$ close to the radial modes when they behave
as acoustic modes, while $E$ is somewhat higher where the modes have
substantial g-mode character.
As is clear from the figure the two modes exchange character at the
avoided crossing; 
at the closest approach, near the vertical line, $E$ is the same in the
two modes.

The corresponding results for $l = 2$ are shown in Fig.~\ref{fig:eboofreq2}.
Here the number of g-mode branches is higher.
Again, these undergo avoided crossings with the acoustic modes,
but the minimum separation is so small that the avoided nature is
barely discernible.
Also, the mode inertias, shown in panel (b) of the figure,
on the g-mode branches exceed those of the radial modes by almost
two orders of magnitude (note the logarithmic scale).
The reason for this difference in behaviour between $l = 1$ and 2
is the greater extent of the evanescent region in the latter case
(cf.\ Fig.~\ref{fig:eboochar}).
This decreases the coupling between the g-mode and acoustic regions,
making the g modes much more strongly trapped in the innermost
parts of the model, and decreasing the interval over which the
avoided crossing take place \citep[see also][]{Christ1980a}.

Oscillations with g-mode character are much more sensitive to the properties
of stellar cores than are the purely acoustic modes. 
In particular, Eq.~(\ref{eq:buoy}) shows that the buoyancy frequency
depends directly on the gradient in the hydrogen abundance.
Thus such {\it mixed modes} have a very substantial asteroseismic
diagnostic potential.
An important issue is whether the modes can be expected to be excited
to observable amplitudes.
It is perhaps not unreasonable to assume that the stochastic excitation excites
the modes to a total energy that depends predominantly on the properties
of the oscillations in the near-surface layers, where both the excitation and
the dominant damping take place.
In these superficial layers the characteristic timescale of the convection
is of the same order of magnitude as the pulsation period of the
oscillations \citep[e.g.][]{Goldreich77, Christ83}. This leads to a strong 
coupling between the global oscillations and the turbulent velocity field 
and consequently efficient mode excitation by the convection
\citep[e.g.][]{Balmforth92b, Houdek2006}.
Since the properties of the oscillations in this region, for low-degree modes,
are mainly determined by the frequency, we would therefore expect that the
mode energy $\CE = \CE(\nu)$ is also a function of frequency.
Using that $\CE \sim E M \langle V^2 \rangle$ (cf.\ Eq. \ref{eq:inertia}),
where $\langle V^2 \rangle$ is the mean square amplitude,
we might therefore expect 
\begin{equation}
\langle V^2 \rangle^{1/2} \propto E^{-1/2} \; ,
\label{eq:ampl}
\end{equation}
at a given frequency.
Based on Figs~\ref{fig:eboofreq} and \ref{fig:eboofreq2} we would
therefore predict that for $l = 1$ even modes on the g-mode branch
would have amplitudes that could be observable, whereas for $l = 2$
modes with some g-mode character would only be visible in the narrow
regions around the avoided crossings, for this model.%
\footnote{A more careful analysis of the visibility of the modes
in terms of the peak height in the power spectra, rather than the
root-mean-square velocity, shows a somewhat more complex picture,
with strong dependence on the mode lifetime in relation to the duration 
of the observations; see \citet{Fletch2006}, \citet{Dupret2009}.}

\begin{figure}[t]
\centering
\includegraphics[width=0.45\textwidth]{\fig/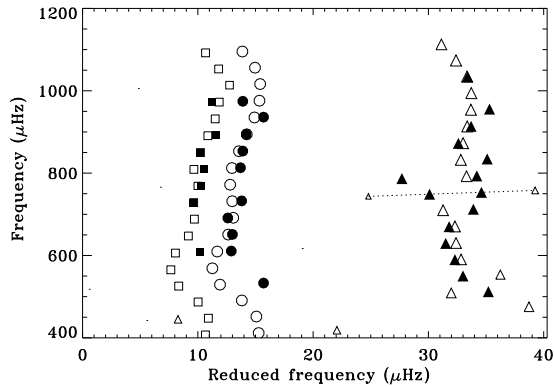}
\caption{%
\'Echelle diagram (see text),
with a starting frequency of $33.7 \muHz$ and a large separation 
$\Delta \nu = 40.3 \muHz$.
Modes of degree $l = 0$, 1 and 2 are shown with circles, triangles and squares,
respectively.
Open symbols show results for the best-fitting model of $\eta$ Boo
\citep{DiMaur2003},
marked by a vertical line in Figs~\ref{fig:eboofreq} and \ref{fig:eboofreq2},
while filled symbols are for combined observations of
\citet{Kjelds2003} and \citet{Carrie2005}.
\label{fig:ebooech}
} 
\end{figure}

It is of great interest to compare these model properties with the
observed frequencies.
We base this on a re-analysis by H.\ Kjeldsen (unpublished) of the
observations by \citet{Kjelds2003} and \citet{Carrie2005} 
\citep[see][for details]{Aerts2009}.
The results are illustrated in an {\it \'echelle diagram} 
\citep[see][]{Grec1983} in Fig.~\ref{fig:ebooech}.
Here the frequency spectrum has been divided into slices of a 
length given by the large separation $\Delta \nu$, starting at
a frequency of $33.7 \muHz$.
These reduced frequencies are plotted against the starting frequency
of each slice, corresponding, in the figure, essentially to
stacking the slices.
According to the asymptotic expression for acoustic modes,
Eq.~(\ref{eq:asymptotics}), one would expect the result to be
nearly vertical sequences of points, separated by the small separation,
although perhaps with some variation arising from the acoustic glitch
associated with the second helium ionization zone
(cf.\ Section~\ref{sec:glitch}).
For the observations, shown by filled symbols,
this is indeed satisfied by the frequencies for $l = 0$ and 2.
However, the points for $l = 1$ clearly show a less regular behaviour.

To interpret these results the open symbols show corresponding results
for a model matched to the observed frequencies, as well as the
effective temperature and luminosity of the star;
this model corresponds to the vertical lines in 
Figs~\ref{fig:eboofreq} and \ref{fig:eboofreq2}.
The frequencies were corrected for near-surface effects
according to the procedure of \citet{Kjelds2008} 
(see Section~\ref{sec:surfprob}).
To provide some indication of the visibility of the modes, 
the size of the symbols has been scaled to reflect the 
amplitude estimate in Eq.~(\ref{eq:ampl}), relative to a radial
mode of the same frequency.
For $l = 2$ this essentially makes the modes on the g-mode 
branches invisible, owing to their large $E$.
However, the $l = 1$ modes are retained, with a behaviour showing
strong departures from the acoustic asymptotic relation, for
modes with a strong mixed nature.
As an example, the dotted line connects the two modes shown to 
undergo an avoided crossing in Fig.~\ref{fig:eboofreq};
as expected for modes at the closest point of an avoided crossing
these are distributed equally on either side of the expected
location of the acoustic mode.
Although there is no detailed agreement between the observed
and computed frequencies of the $l = 1$ modes, the strong
resemblance between their behaviour suggests that the observations
show evidence for mixed modes in this star.

\subsection{Superficial problems}

\label{sec:surfprob}
%
A major issue in the analysis of solar-like oscillation frequencies is
the influence of the near-surface layers. 
Stellar modelling normally treats these layers in a highly simplified
manner, with a simple model of the stellar atmosphere and a simplified 
treatment of the upper, superadiabatic,
parts of the convection zone, such as the mixing-length formulation
\citep{Bohm1958}.
Also, the effects of the turbulent Reynolds stresses on the structure are
almost always ignored.
In computations of oscillation frequencies the adiabatic approximation is
commonly adopted, and perturbations to the turbulent 
Reynolds stresses are neglected.
When nonadiabatic calculations are carried out the treatment of the
perturbation to the convective flux is a major uncertainty in the calculation
\citep[e.g.,][]{BakerGough79, Gough80, Balmforth92a}.

These problems are confined to the stellar atmosphere and the outermost part 
of the the stellar interior.
In these regions the properties of modes of low or intermediate degree are
largely independent of degree, at least in spherically symmetric models, and
hence the influence of the near-surface problems on the oscillation frequency
is essentially just a function of frequency \citep[e.g.,][]{Christ1997};
for modes of intermediate degree there is some dependence on degree in the 
form of a trivial scaling with mode inertia.
In the solar case, where modes over a broad range of degree are observed,
these properties allow the near-surface errors in the model to be isolated and
suppressed in the analysis of the observed frequencies.
In fact, for models such as Model S they dominate the differences between the
solar and model frequencies \citep[e.g.,][]{Christ1996}.
This provides an estimate of the effects in the solar case, showing that they 
are very small at low frequency and increase rapidly with frequency, as 
indeed expected from simple analyses \citep[e.g.,][]{Christ1980b}.
As a consequence, they also have a substantial effect on the determination
of the large frequency separation (see Eq.~\ref{eq:largesep}).

Such a separation is not possible for stellar observations, restricted to low
degree;
hence the near-surface effects must be kept in mind as a source of systematic
error in the analysis.
As discussed in Section~\ref{sec:inphase}, the effects are strongly suppressed
in separation ratios (cf.\ Eq.~\ref{eq:seprat}), which are essentially
determined just by the inner phase shift.
This is extremely valuable for the diagnostics of stellar cores.
To obtain further information from the observed frequencies, additional 
assumptions, so far with limited justification, must be made.
The behaviour of the oscillations in the stellar atmosphere is largely
controlled by the value of the frequency, measured in terms of the atmospheric
acoustical cut-off frequency $\omega_{\rm ac}$ which, in the approximation of
an isothermal atmosphere, is given by
\begin{equation}
\omega_{\rm ac} = \frac{c}{2H} \; ,
\end{equation}
where $H$ is the pressure scale height (note that in an isothermal
atmosphere with constant mean molecular weight the pressure scale height
is equal to the density scale height).
Thus, for stars similar to the Sun one can perhaps use the {\it ansatz}
that the near-surface effects can be approximated by
$a f_{\rm surf}^{(\odot)}(\omega / \omega_{\rm ac})$ where 
$f_{\rm surf}^{(\odot)}$ is obtained from the solar observations and
$a$ is a constant to be determined as part of the fit of the model to the
observed frequencies.
In a similar spirit \cite{Kjelds2008} suggested to represent the 
near-surface frequency shift as $a (\omega / \omega_0)^b$, where 
$\omega_0$ is a suitable reference frequency, the exponent $b$ is obtained
from the solar data, and $a$ is obtained from the frequency fit, together 
with a scaling factor that accounts for the difference in mean density between
the star and the model.

It is obvious that these simple procedures have relatively weak justification,
particularly for stars that are somewhat different from the Sun.
Analysis of the detailed data for a broad range of stars that are being 
obtained by the Kepler mission will certainly contribute to a better 
characterization of the near-surface effects.
Also, an overview of how these effects influence the frequencies
of stellar models, depending on the stellar parameters, would be
very interesting.

A key goal is clearly to improve the treatment of the physics of 
the near-surface regions.
In the solar case, models of the near-surface region based on hydrodynamical
simulations result in frequencies in better, but still not complete, 
agreement with the observations \citep{Rosent1999, Li2002}.
Such calculations were also applied with apparent success to the analysis
of MOST observations of $\eta$ Boo by \citet{Straka2006}.
The effects of convection dynamics on the oscillation frequencies were also
discussed by \citet{Balmforth92b}, \citet{Rosenthal95} and \citet{Houdek96}, 
who used the nonlocal, time-dependent formulation by \citet{Gough77a, Gough77b} 
in their linear stability analyses.
These authors reported that the inclusion of the Reynolds stress in an 
equilibrium solar model reduces substantially the frequency difference between
observed and adiabatically computed radial eigenfrequencies.
This difference, however,
becomes larger again when nonadiabaticity and convection dynamic effects
are included in the stability computations,
thereby compensating nearly the
frequency shifts arising from the Reynolds stress in the equilibrium model.
A similar conclusion was reported by \citet{Christ95} for $\eta$ Boo.
The effect of such modelling on other stars should obviously be investigated.

\section{The observational situation}

\label{sec:obs}
%
The tiny amplitudes of solar-like oscillations have greatly complicated
their observation in distant stars, despite extensive efforts.
It was only the observations of \citet{Kjelds1995} that led to
the first detection of individual modes of oscillations in a solar-like star,
the star $\eta$ Boo; this was later confirmed by 
\citet{Kjelds2003} and \citet{Carrie2005}.
However, in the last decade the development in observational techniques
to measure Doppler velocity,
largely motivated by the search for extra-solar planetary systems, has 
led to fairly detailed observations of solar-like oscillations in a number
of stars \citep[for a brief review, see][]{Beddin2008},
reaching a noise level in the amplitude spectrum as low as
$1.4 \, {\rm cm \, s^{-1}}$ \citep{Kjelds2005}.
From space, photometric observations allow simultaneous observations
of large numbers of stars.
The CoRoT mission \citep{Baglin2009} has produced remarkable 
asteroseismic results \citep[e.g.,][]{Michel2008a, Michel2008b},
including striking results on solar-like pulsations in red giants
\citep[e.g.,][]{DeRidd2009}.
Also, the first results from the Kepler mission \citep{Boruck2009} confirm
the great potential of the mission for asteroseismology
\citep[e.g.,][]{Christ2008}.

The observations generally provide a reliable measure of the large frequency
separation (see Eq.\ \ref{eq:largesep}).
To move beyond this an identification of at least the degree of the modes
is required.
In the case, for example, of the binary system $\alpha$ Cen A and B
\citep{Bouchy2001, Bouchy2002, Butler2004, Beddin2004, Bazot2007,
Carrie2003, Kjelds2005}
or for $\beta$ Hydri \citep{Beddin2007} an unambiguous identification has been
possible, based on a match to the asymptotic behaviour 
(cf.\ Eq.~\ref{eq:asymptotics}).
This has led to detailed asteroseismic analyses for these stars, yielding
precise determinations of their overall parameters and some indications of
possible problems in the modelling
\citep[e.g.,][]{Eggenb2004, Miglio2005}.


In other cases, however, the mode identification has proved to be difficult.
An interesting example is the star Procyon~A ($\alpha$ CMi),
of spectral type F5IV-V,
which has been the target of several asteroseismic observing campaigns.
Evidence for a power excess associated with solar-like oscillations was
found already by \citet{Brown1991};
however, several subsequent observations have failed to make an 
unambiguous identification of the modes of oscillation,
and doubt was even raised on the reality of the oscillations on the
basis of observations from the MOST satellite \citep{Matthe2004}
\citep[see, however,][]{Beddin2005}.
A very extensive observing campaign, involving 11 telescopes,
was carried out by \citet{Arento2008}.
This confirmed beyond any doubt that Procyon exhibited solar-like oscillations;
the mode identification has proven to be difficult but is now close to
being completed (Bedding et al., in preparation).
Similar problems have been found in the analysis of data from the CoRoT
satellite, such as the star HD49933, of spectral type F2V
\citep{Appour2008, Benoma2009, Gaulme2009}.

%

These problems in the data analysis appear to a large extent to be associated
with the mode lifetime in these stars, which is substantially shorter than
in, for example, the Sun or $\alpha$ Cen A and B, hence leading to a more
complex spectrum of oscillations.
Thus it is of substantial interest that a close relation has been found 
between the effective temperature $T_{\rm eff}$ of low-mass main-sequences 
stars and the average mode lifetime $\langle\tau\rangle$ of the five most
prominent modes:
\begin{equation}
\langle\tau\rangle\propto T^{-4}_{\rm eff}
\end{equation}
\citep{Chapli2009}.
This relation can obviously be of importance in the simulation of 
asteroseismic oscillations and the selection of targets.
However, it should in any case not be a surprise that further development
of the analysis techniques is required for stars whose properties differ
substantially from those of the Sun.
These developments will of course be inspired by the extensive data 
obtained from CoRoT and expected from Kepler.
However, analysis of realistically simulated data for a variety of
stellar parameters is also very valuable, ideally carried out in blind,
so-called `Hare and Hounds', investigations.
Tools have been developed for such simulations \citep[e.g.,][]{DeRidd2006},
and they are being used in extensive analyses in the asteroFLAG project,
with specific application to the data expected from the Kepler mission
\citep{Chapli2008}.


\section{Effects of model physics}

%
An ultimate goal of helio- and asteroseismology is obviously to relate
the inferences to the physics of stellar interiors and hence to get 
a better understanding of the processes taking place in stars.
Here we consider a few relevant examples, including some that were
highlighted at the workshop, and consider effects on stellar structure that
are, or may be, amenable to helio- and asteroseismic investigations.

\begin{figure}[t]
\centering
\includegraphics[width=0.45\textwidth]{\fig/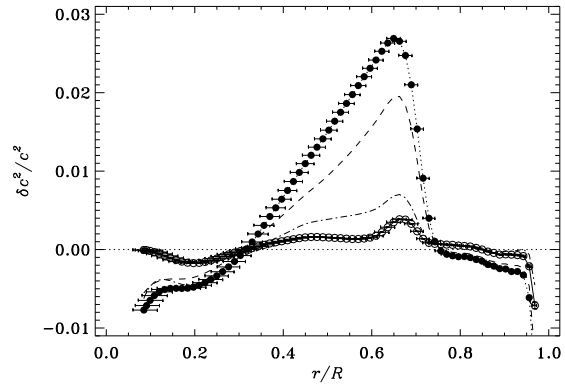}
\caption{%
Relative differences in squared sound speed between the Sun and three solar 
models, in the sense (Sun) -- (model), inferred from inversion of the 
so-called `Best set' of observed frequencies \citep[see][]{Basu1997}.
The barely visible vertical bars mark $1\,\sigma$ errors in the inferences,
while the horizontal bars indicate the resolution of the inversion.
Open symbols: Model S of \citet{Christ1996};
filled circles: corresponding model, but assuming the
\citet{Asplun2005a} composition; dashed curve: corresponding model,
but assuming the \citet{Asplun2009} composition;
dot-dashed curve: model corresponding to Model S, but switching off
electron screening (in the core this can hardly be distinguished from 
the dashed curve)
\label{fig:csqinv}
} 
\end{figure}

\subsection{The solar sound speed}

\label{sec:soundspeed}
%
The observed solar oscillation frequencies cover a broad range in
spherical-harmonic degree.
This has allowed inverse analyses to determine solar internal properties,
in particular the sound speed, with high accuracy in most of the solar interior.
Solar models computed a decade ago generally provided relatively good agreement
with these inferences, with maximum relative deviations 
in the square of the sound speed well below 1~per cent.
A typical example is illustrated by the open symbols in Fig.~\ref{fig:csqinv},
based on the so-called Model S of \citet{Christ1996};
in common with many solar models of the epoch it assumed the 
\citet{Greves1993} solar composition.
However, starting with \citet{Allend2001}
reanalyses of solar spectral lines have led to substantial revisions in
the determination of the surface abundances of, in particular,
carbon, nitrogen and oxygen, leading to a reduction in the ratio 
$\Zs/\Xs$ between the abundances by mass $\Zs$ of heavy elements and
$\Xs$ of hydrogen from 0.0245 to 0.0165.
This revision resulted from the use of hydrodynamical models of the
solar atmosphere and the inclusion of NLTE effects in the analysis 
\citep[for reviews, see][]{Asplun2005a, Asplun2005b}.
The main effect on solar modelling is a corresponding reduction in the opacity,
leading to a substantial change in the sound speed in the radiative interior.
The effect on the sound speed is illustrated by the filled symbols in
Fig.~\ref{fig:csqinv}.%
\footnote{Here, and in the following results on solar modelling,
the models correspond to Model S except in the specific property 
under investigation; the models have all been calibrated to solar
radius, luminosity and the assumed value of $\Zs/\Xs$ at solar age,
by adjusting the mixing length and the initial composition.}
It is evident that this has led to a serious deterioration in the agreement
between the solar models and the helioseismic inference.

\begin{figure}[t]
\centering
\includegraphics[width=0.45\textwidth]{\fig/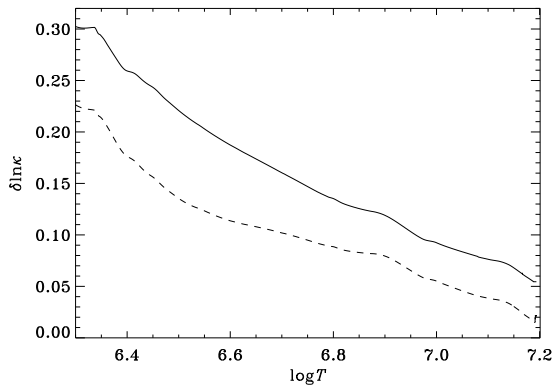}
\caption{%
Changes to the natural logarithm $\ln \kappa$ of the opacity, required to 
bring models with revised composition in accordance with Model S of
\citet{Christ1996}.
Solid curve: correction for the \citet{Asplun2005a} composition;
dashed curve: correction for the \citet{Asplun2009} composition.
\citep[See][]{Christ2009}
\label{fig:dlkap}
} 
\end{figure}

This conflict between helioseismology and the solar models resulting from
the revised composition was discussed in detail by \citet{Basu2008};
possible solutions were reviewed by \citet{Guzik2006}, with the conclusion
that none of the proposals were fully satisfactory.
Since the dominant effect on the models of the change in the composition
arises from the resulting change in opacity,
a simple solution would be to make corresponding changes to the
intrinsic opacity \citep[e.g.,][]{Bahcal2005}.
\citet{Christ2009} estimated the opacity change required to restore the
model computed with the old composition;
the result is illustrated in Fig.~\ref{fig:dlkap}.
It is perhaps questionable if such a major change, of up to 30 per cent,
is physically justifiable, although there are undoubtedly significant
uncertainties in the opacity calculations.

Recently, \citet{Asplun2009} reconsidered the abundance determinations,
using improved solar atmosphere models and a careful selection of
atomic data and spectral lines
\citep[see also][this volume]{Greves2010}.
This resulted in a modest increase in the inferred abundances, such that
$\Zs/\Xs = 0.0181$.
The effect on the comparison between the resulting solar model and
helioseismic inferences is illustrated by the dashed line in
Fig.~\ref{fig:csqinv}.
Obviously the model is somewhat closer to the solar sound speed although
the differences are still much larger than for the original model.
The opacity change required to restore the original model is illustrated
in Fig.~\ref{fig:dlkap}; the maximal change, around 23 per cent, may still
be larger than can reasonably be explained by errors in the opacity
determinations.

%
Other uncertainties in stellar modelling may have a noticeable effect on
the comparison with helioseismic inferences.
An interesting example is the effect of electron screening on the
nuclear reactions.
This is normally treated in the Debye-H\"uckel approximation,
according to a formulation developed by \citet{Salpet1954};
here the formation of clouds of negative charge around the reacting
nuclei leads to a reduction in the Coulomb repulsion and hence an
increase in the reaction rates.
However, it has been pointed out 
that the underlying mean-field approximation may be invalid under
stellar conditions
\citep[see, for example,][for a review]{Shaviv2004}.
As discussed by \citet[][this volume]{Mussac2010}
\citep[see also][]{Mao2009} 
this has motivated calculations of the properties of stellar plasmas
using the techniques of molecular dynamics.
The results indicate that the effects of electron screening are weaker
than in the Salpeter formulation and that the interactions between the
charged particles may in fact lead to a {\it reduction} in the
reaction rates.
To investigate the consequences for solar modelling we have computed a
model where electron screening was simply switched off.
The resulting sound-speed difference is illustrated by the 
dot-dashed line in Fig.~\ref{fig:csqinv}.
Although less dramatic than the effect of the change in composition,
the change in the nuclear reactions,
and the resulting change in the composition required to calibrate the model,
leads to a definite deterioration
in the agreement with the helioseismically inferred sound speed;
such an effect of suppressing electron screening
was already pointed out by \citet{Weiss2001}.

\subsection{The onset of convective cores}

\label{sec:concore}
%
A major uncertainty in the modelling of intermediate- and high-mass stars
is the treatment of convective cores.
Convective cores are typically found in stars of masses just a little 
higher than the mass of the Sun, thus including also stars where
solar-like oscillations are expected.
As discussed in Section~\ref{sec:inphase} the detailed properties of
the core are reflected in the small frequency separations and hence
there are excellent prospects to use asteroseismology to constrain
the modelling of convective cores.
A detailed analysis of this was carried out by 
\citet{Popiel2005}, emphasizing the effect of the composition 
discontinuity caused by the growing convective core in stars of mass
up to around $1.9 \Msun$.


Here we consider the effects 
of modifications to the microphysics and composition of the stellar interiors.
The convective instability is caused by the increasing temperature sensitivity
of nuclear reactions as the CNO cycle, with increasing stellar mass,
becomes dominant in hydrogen burning.
An additional contribution is the gradual conversion of
${}^{16}{\rm O}$ to ${}^{14}{\rm N}$ through a branch of the CNO cycle;
this increases the abundance of ${}^{14}{\rm N}$ which controls the
overall rate of the energy generation from the CNO cycle and hence its
importance, leading to the growth of the mass of the convective core.

\begin{figure}[t]
\centering
\includegraphics[width=0.45\textwidth]{\fig/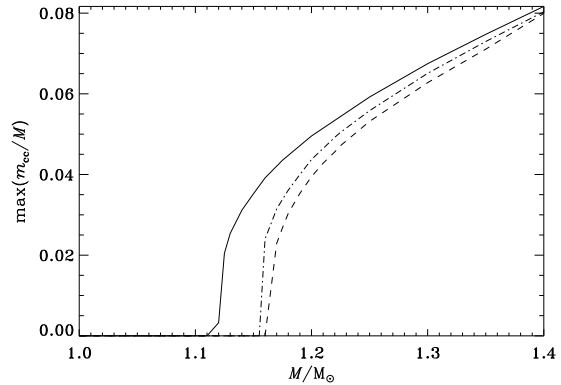}
\caption{%
Maximum fractional mass of the convective core during main-sequence evolution,
as a function stellar mass in units of the solar mass.
The models used the \citet{Angulo1999}
nuclear parameters and included electron screening;
diffusion and settling was not included.
Solid line: reference model, with the \citet{Greves1993} composition,
and initial hydrogen and heavy-element abundances given by
$(X_0, Z_0) = (0.708, 0.0196)$;
dashed line: \citet{Asplun2005a} composition,
and $(X_0, Z_0) = (0.730, 0.0137)$, 
dot-dashed line: \citet{Asplun2009} composition,
and $(X_0, Z_0) = (0.722, 0.0149)$
\label{fig:agscore}
} 
\end{figure}

It is obvious that the CNO cycle is sensitive to the abundances of 
carbon, nitrogen and oxygen.
Abundances of solar-like stars are typically determined relative
to the solar abundance, and hence the recent revisions of the
solar abundances have a similar effect on the assumed abundances of
stars and hence potentially on the onset of convective cores.
In fact, \citet{Vanden2007} showed that the \citet{Asplun2005b}
abundances led to isochrones for the open cluster M67 which did not
show a `hook', resulting from the presence of a convective core.
This was in contrast to models with the old composition, which showed
a clear hook, and in apparent contradiction to observed colour-magnitude
diagrams for the cluster.
To investigate this further, we have considered the properties of
convective cores as a function of stellar mass,
characterizing them by the maximum mass reached by the core during 
central hydrogen burning.
The models neglected diffusion and settling and used the NACRE
\citep{Angulo1999} nuclear parameters, but otherwise largely 
corresponded to Model S of \citet{Christ1996}.
The results are illustrated in Fig.~\ref{fig:agscore}.
In agreement with \citet{Vanden2007} the mass at onset of the convective 
core increases by around $0.04 \Msun$ as a result of the
\citet{Asplun2005b} revision of the composition.
The modest revision by \citet{Asplun2009} has only a slight effect,
in comparison.

\begin{figure}[t]
\centering
\includegraphics[width=0.45\textwidth]{\fig/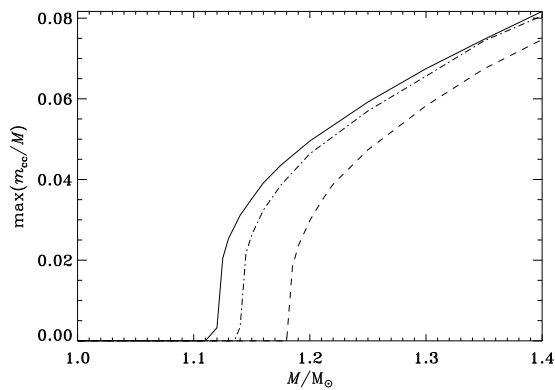}
\caption{%
As Fig.~\ref{fig:agscore}, all models having the \citet{Greves1993}
composition.
Solid line: reference model;
dashed line: reduced ${}^{14}{\rm N}$ reaction rate \citep{Angulo2005};
dot-dashed line: no electron screening
\label{fig:n14core}
} 
\end{figure}

The presence of convective cores is obviously also sensitive to nuclear
properties, particularly for the reaction 
${}^{14}{\rm N}({}^1{\rm H}, \gamma){}^{15}{\rm O}$ which controls 
the rate of the dominant part of the CNO cycles.
From new laboratory experiments and analyses,
\citet{Angulo2005} found a large reduction in this rate.
As shown in Fig.~\ref{fig:n14core} this has a substantial effect on
the minimum mass of stars with convective cores,
increasing it by $0.06 \Msun$.
On the other hand, since the CNO cycle makes a modest contribution to
solar energy generation the effect on solar models is largely 
insignificant.

We have also considered the effect on the convective cores of switching off
electron screening; 
the effect of electron screening increases with the charges of the 
interacting nuclei and hence turning it off decreases the relative 
importance of the CNO cycle, again delaying the onset of convective cores.
This is also illustrated in Fig.~\ref{fig:n14core}; the effect is
relatively modest.

Further investigations will be required to test the effects 
on the oscillation frequencies of these changes in the model properties,
compared with other uncertainties in the treatment of convective cores,
and the prospects for applying asteroseismic diagnostics, such
as those proposed by \citet{Cunha2007}, to investigate them.

\section{Prospects for the future}

Asteroseismology is being revolutionized in the present period.
As documented in a recently published special Volume 506 of
{\it Astronomy and Astrophysics}, the initial results from the CoRoT mission
are very promising, and at the time of writing the very early data from
the Kepler mission are being analysed, through an intense and coordinated
effort in the Kepler Asteroseismic Science Consortium, with a view towards
publication in early 2010.
Given the continuing operations of CoRoT and the very extensive data to be
obtained with Kepler it is clear that coming years will see a huge effort in
the analysis of asteroseismic data, with a corresponding increase in our
information on and, one may hope, understanding of stellar properties.
On a somewhat longer timescale there is hope that the PLATO mission
\citep{Catala2008} will provide detailed asteroseismic data on 
tens of thousands of stars.

Despite the impressive success of space-based asteroseismic investigations,
there is still a need for ground-based observations.
As found early in observations of the Sun \citep{Harvey1988} the
intrinsic stellar noise is much higher, relative to the oscillations,
in photometric observations than in observations of Doppler velocity.
Velocity observations can be carried out from the ground, the main limiting
factor being the access to the required instrumentation, at several
sites to ensure continuous data and over a sufficient length of time.
This motivates the establishment of dedicated facilities;
an example is the Stellar Observations Network Group (SONG) project
\citep{Grunda2009}, aiming to establish a network of 8 observatories,
with a suitable world-wide distribution, equipped with 1\,m telescopes
and high-resolution spectrographs to carry out asteroseismic Doppler-velocity 
observations.
The proposal to set up the SIAMOIS facility for Doppler-velocity 
observations on Dome C in Antarctica \citep{Mosser2008} is also
very promising.

In parallel with these great observational strides there is a strong
need for further development of the techniques for stellar modelling and
the interpretation of the asteroseismic data.
It is already clear that, not surprisingly, the oscillation properties of
other stars show striking departures from the familiar properties of
solar oscillations and that new techniques will be needed to optimize
the analysis of the data and the use of the results for asteroseismic
inferences.
From the results of these analyses we can surely expect further surprises
and identification of deficiencies in our stellar modelling,
leading to new insights into the physics of stellar interiors.

\acknowledgments
We are very grateful to Maria Pia Di Mauro for the organization of 
a very interesting workshop, and for a long and warm friendship.
This work was supported by the Danish Natural Science Research Council
and by the European Helio- and Asteroseismology Network (HELAS),
a major international collaboration funded by the European Commission's
Sixth Framework Programme.
GH acknowledges support by the Austrian Science Fund (FWF project P21205).

\nocite{*}
\bibliographystyle{spr-mp-nameyear-cnd}
\bibliography{biblio-u1}

\begin{thebibliography}{}

\bibitem[Aerts {et~al.}(2009)]{Aerts2009}
Aerts, C., Christensen-Dalsgaard, J., Kurtz, D.~W.:
{\it Asteroseismology},
Springer, Heidelberg
(2009, in the press)

\bibitem[Allende Prieto {et~al.}(2001)]{Allend2001}
Allende Prieto, C., Lambert, D.~L., Asplund, M.:
{\apj} {\bf 556}, L63 
(2001)

\bibitem[Angulo {et~al.}(1999)]{Angulo1999}
Angulo, C., Arnould, M., Rayet, M., et~al.:
{\nphysa} {\bf 656}, 3 
(1999)

\bibitem[Angulo {et~al.}(2005)]{Angulo2005}
Angulo, C., Champagne, A.~E., Trautvetter, H.-P.:
{\nphysa} {\bf 758}, 391c 
(2005)

\bibitem[Appourchaux {et~al.}(2008)]{Appour2008}
Appourchaux, T., Michel, E., Auvergne, M., et al.:
{\aap} {\bf 488}, 705 
(2008)

\bibitem[Arentoft {et~al.}(2008)]{Arento2008}
Arentoft, T., Kjeldsen, H., Bedding, T.~R., et al.:
{\apj} {\bf 687}, 1180 
(2008)

\bibitem[Asplund(2005)]{Asplun2005a}
Asplund, M.:
{\araa} {\bf 43}, 481 
(2005)

\bibitem[Asplund {et~al.}(2005)]{Asplun2005b}
Asplund, M., Grevesse, N., Sauval, A.~J.:
In {Cosmic Abundances as Records of Stellar Evolution and Nucleosynthesis},
eds T.~G.\ Barnes III, F.~N.\ Bash, ASP Conf. Ser. {\bf 336},
p.\,25 
(2005)

\bibitem[Asplund {et~al.}(2009)]{Asplun2009}
Asplund, M., Grevesse, N., Sauval, A.~J., Scott, P.:
{\araa} {\bf 47}, 481 
(2009)

\bibitem[Baglin {et~al.}(2009)]{Baglin2009}
Baglin, A., Auvergne, M., Barge, P., Deleuil, M., Michel, E.\ and
the CoRoT Exoplanet Science Team:
In {Proc. IAU Symp. 253, Transiting Planets},
eds F.\ Pont, D.\ Sasselov, M.\ Holman,
IAU and Cambridge Univ. Press, p.\,71 
(2009)

\bibitem[Bahcall {et~al.}(2005)]{Bahcal2005}
Bahcall, J.~N., Basu, S., Pinsonneault, M., Serenelli, A.~M.:
{\apj} {\bf 618}, 1049 
(2005)

\bibitem[Baker \& Gough(1979)]{BakerGough79}
Baker N., Gough D.~O.:
\apj\ \textbf{234}, 232
(1979)

\bibitem[Balmforth(1992a)]{Balmforth92a}
Balmforth N.~J.:
\mnras\ \textbf{255}, 603
(1992a)

\bibitem[Balmforth(1992b)]{Balmforth92b}
Balmforth N.~J.:
\mnras\ \textbf{255}, 632
(1992b)

\bibitem[Basu \& Antia(2004)]{Basu2004}
Basu, S., Antia, H.~M.:
{\apj} {\bf 606}, L85 
(2004)

\bibitem[Basu \& Antia(2008)]{Basu2008}
Basu, S., Antia, H.~M.:
{\physrep} {\bf 457}, 217 
(2008)

\bibitem[Basu {et~al.}(1994)]{Basu1994}
Basu S., Antia H.~M., Narasimha D.:
\mnras\ 267, 209
(1994)

\bibitem[Basu {et~al.}(1997)]{Basu1997}
Basu, S., Chaplin, W.~J., Christensen-Dalsgaard, J.,
Elsworth, Y., Isaak, G.~R., New, R., Schou, J.,
Thompson, M.~J., Tomczyk, S.:
{\mnras} {\bf 292}, 243 
(1997)

\bibitem[Basu {et~al.}(2002)]{Basu2002}
Basu, S., Christensen-Dalsgaard, J., Thompson, M.~J.:
In {Proc. 1st Eddington Workshop, `Stellar Structure
and Habitable Planet Finding'},
eds F.\ Favata, I.~W.\ Roxburgh and D.\ Galad\'{\i}-Enr\'{\i}quez,
ESA SP-485, ESA Publications Division, Noordwijk, The Netherlands,
p.\,249 
(2002)

\bibitem[Basu {et~al.}(2007)]{Basu2007}
Basu S., Chaplin W.~J., Elsworth Y., New A.~M., Serenelli G., Verner G.~A.:
\apj, {\bf 655}, 660 
(2007)

\bibitem[Bazot {et~al.}(2007)]{Bazot2007}
Bazot, M., Bouchy, F., Kjeldsen, H., Charpinet, S., Laymand, M.,
Vauclair, S.:
{\aap} {\bf 470}, 295 
(2007)

\bibitem[Bedding \& Kjeldsen(2008)]{Beddin2008}
Bedding, T.~R., Kjeldsen, H.:
In {Proc. 14$^{th}$ Cambridge Workshop on Cool Stars, Stellar Systems, and
the Sun}, ed.\ G.~T.\ van Belle, {ASP Conf. Ser.} {\bf 384}, p.\,21 
(2008)

\bibitem[Bedding {et~al.}(2007)]{Beddin2007}
Bedding, T.~R., Kjeldsen, H., Arentoft, T., et al.:
{\apj} {\bf 663}, 1315 
(2007)

\bibitem[Bedding {et~al.}(2005)]{Beddin2005}
Bedding, T.~R., Kjeldsen, H., Bouchy, F.,
Bruntt, H., Butler, R.~P., Buzasi, D.~L., Christensen-Dalsgaard, J.,
Frandsen, S., Lebrun, J.-C., Marti\'c, M., Schou. J.:
{\aap} {\bf 432}, L43 
(2005)

\bibitem[Bedding {et~al.}(2004)]{Beddin2004}
Bedding, T.~R., Kjeldsen, H., Butler, R.~P., McCarthy, C., 
Marcy, G.~W., O'Toole, S.~J., Tinney, C.~G., Wright, J.~T.:
{\apj} {\bf 614}, 380 
(2004)

\bibitem[Benomar {et~al.}(2009)]{Benoma2009}
Benomar, O., Appourchaux, T., Baudin, F.:
{\aap} {\bf 506}, 15 
(2009)

\bibitem[B\"ohm-Vitense(1958)]{Bohm1958}
B\"ohm-Vitense, E.:
{Z.\ Astrophys.} {\bf 46}, 108 
(1958)

\bibitem[Borucki {et~al.}(2009)]{Boruck2009}
Borucki, W., Koch, D., Bathalha, N., et al.:
In {Proc. IAU Symp. 253, Transiting Planets},
eds F.\ Pont, D.\ Sasselov, M.\ Holman,
IAU and Cambridge Univ. Press, p.\,289 
(2009)

\bibitem[Bouchy \& Carrier(2001)]{Bouchy2001}
Bouchy, F., Carrier, F.:
{\aap} {\bf 374}, L5 
(2001)

\bibitem[Bouchy \& Carrier(2002)]{Bouchy2002}
Bouchy, F., Carrier, F.:
{\aap} {\bf 390}, 205 
(2002)

\bibitem[Brown {et~al.}(1991)]{Brown1991}
Brown, T.~M., Gilliland, R.~L., Noyes, R.~W., Ramsey, L.~W.:
{\apj} {\bf 368}, 599 
(1991)

\bibitem[Bruntt(2007)]{Bruntt2007}
Bruntt, H.:
{\rm Comm. in Asteroseismology} {\bf 150}, 326 
(2007)

\bibitem[Butler {et~al.}(2004)]{Butler2004}
Butler, R.~P., Bedding, T.~R., Kjeldsen, H., McCarthy, C., O'Toole, S.~J.,
Tinney, C.~G., Marcy, G.~W., Wright, J.~T.:
{\apj} {\bf 600}, L75 
(2004)

\bibitem[Buzasi {et~al.}(2005)]{Buzasi2005}
Buzasi, D. L., Bruntt, H., Bedding, T. R., et al.:
{\apj} {\bf 619}, 1072 
(2005)

\bibitem[Carrier \& Bourban(2003)]{Carrie2003}
Carrier, F., Bourban, G.:
{\aap} {\bf 406}, L23 
(2003)

\bibitem[Carrier {et~al.}(2005)]{Carrie2005}
Carrier, F., Eggenberger, P., Bouchy, F.:
{\aap} {\bf 434}, 1085 
(2005)

\bibitem[Catala(2008)]{Catala2008}
Catala, C., and the PLATO consortium:
In {Proc. HELAS II International Conference: Helioseismology,
Asteroseismology and the MHD Connections},
eds L.\ Gizon, M.\ Roth,
{J.\ Phys.: Conf. Ser.} {\bf 118}, 012040 
(2008)

\bibitem[Chaplin {et~al.}(2008)]{Chapli2008}
Chaplin, W.~J., Appourchaux, T., Arentoft, T., et al.:
{\rm Astron. Nach.} {\bf 329}, 549 
(2008)

\bibitem[Chaplin {et~al.}(2009)]{Chapli2009}
Chaplin, W.~J., Houdek, G., Karoff, C., Elsworth, Y., New, R.:
{\aap} {\bf 500}, L21 
(2009)

\bibitem[Christensen-Dalsgaard(1980)]{Christ1980a}
Christensen-Dalsgaard, J.:
{\mnras} {\bf 190}, 765 
(1980)

\bibitem[Christensen-Dalsgaard(1984)]{Christ1984}
Christensen-Dalsgaard J.:
In {Proc. Workshop Space research prospects in stellar activity and
variability}, eds A.\ Mangeney, F.\ Praderie, Obs. Paris-Meudon, p.\,11
(1984)

\bibitem[Christensen-Dalsgaard(1993)]{Christ1993}
Christensen-Dalsgaard J.:
In {Proc. GONG 1992: Seismic investigation of the Sun and stars},
ed.\ T.~M.\ Brown, ASP Conf. Ser. {\bf 42}, 
p.\,347
(1993)

\bibitem[Christensen-Dalsgaard(2002)]{Christ2002}
Christensen-Dalsgaard, J.:
{Rev. Mod. Phys.} {\bf 74}, 1073 
(2002)

\bibitem[Christensen-Dalsgaard(2004)]{Christ2004}
Christensen-Dalsgaard, J.:
{\solphys} {\bf 220}, 137 
(2004)

\bibitem[Christensen-Dalsgaard \& Frandsen(1983)]{Christ83}
Christensen-Dalsgaard J., Frandsen S.:
{\solphys} \textbf{82}, 469 
(1983)

\bibitem[Christensen-Dalsgaard \& Gough(1980)]{Christ1980b}
Christensen-Dalsgaard, J., Gough, D.~O.:
{\nat} {\bf 288}, 544 
(1980)

\bibitem[Christensen-Dalsgaard \& P\'erez Hern\'andez(1992)]{Christ1992}
Christensen-Dalsgaard, J., P\'erez Hern\'andez, F.:
{\mnras} {\bf 257}, 62 
(1992)

\bibitem[Christensen-Dalsgaard \& Thompson(1997)]{Christ1997}
Christensen-Dalsgaard, J., Thompson, M.~J.:
{\mnras} {\bf 284}, 527 
(1997)

\bibitem[Christensen-Dalsgaard {et~al.}(1995)]{Christ95}
Christensen-Dalsgaard J., Bedding T., Houdek G., Kjeldsen H.,
Rosenthal C.~S., Trampedach R., Monteiro M.~J.~P.~F.~G., Nordlund, \AA.:
In Proc. IAU Colloq. \textbf{155}, Astrophysical Applications of
Stellar Pulsation, 
eds R.~S.\ Stobie, P.~A.\ Whitelock, PASP~83, p.\,447
(1995)

\bibitem[Christensen-Dalsgaard {et~al.}(1996)]{Christ1996}
Christensen-Dalsgaard, J., D\"appen, W., Ajukov, S.~V., et al.:
{Science} 1286 
(1996)

\bibitem[Christensen-Dalsgaard {et~al.}(2007)]{Christ2008}
Christensen-Dalsgaard, J., Arentoft, T., Brown, T.~M., Gilliland, R.~L.,
Kjeldsen, H., Borucki, W.~J., Koch, D.:
In {Proc. HELAS II International Conference: Helioseismology,
Asteroseismology and the MHD Connections},
eds L.\ Gizon, M.\ Roth,
{J.\ Phys.: Conf. Ser.} {\bf 118}, 012039 
(2008)

\bibitem[Christensen-Dalsgaard {et~al.}(2009)]{Christ2009}
Christensen-Dalsgaard, J., Di Mauro, M.~P., Houdek, G., Pijpers, F.:
{\aap} {\bf 494}, 205 
(2009)

\bibitem[Cunha \& Metcalfe(2007)]{Cunha2007}
Cunha, M.~S., Metcalfe, T.~S.:
{\apj} {\bf 666}, 413 
(2007)

\bibitem[De Ridder {et~al.}(2006)]{DeRidd2006}
De Ridder, J., Arentoft, T., Kjeldsen, H.:
{\mnras} {\bf 365}, 595 
(2006)

\bibitem[De Ridder {et~al.}(2009)]{DeRidd2009}
De Ridder, J., Barban, C., Baudin, F., et al.:
{\nat} {\bf 459}, 398 
(2009)

\bibitem[Di Mauro {et~al.}(2003)]{DiMaur2003}
Di Mauro, M.~P., Christensen-Dalsgaard, J., Kjeldsen, H., Bedding, T.~R.,
Patern\`o, L.:
{\aap} {\bf 404}, 341 
(2003)

\bibitem[Dupret {et~al.}(2009)]{Dupret2009}
Dupret, M.-A., Belkacem, K., Samadi, R., et al.:
{\aap} {\bf 506}, 57 
(2009)

\bibitem[Eggenberger {et~al.}(2004)]{Eggenb2004}
Eggenberger, P., Charbonnel, C., Talon, S., Meynet, G., Maeder, A.,
Carrier, F., Bourban, G.:
{\aap} {\bf 417}, 235 
(2004)

\bibitem[Fletcher {et~al.}(2006)]{Fletch2006}
Fletcher, S.~T., Chaplin, W.~J., Elsworth, Y., Schou, J., Buzasi, D.:
{\mnras} {\bf 371}, 935 
(2006)

\bibitem[Gaulme {et~al.}(2009)]{Gaulme2009}
Gaulme, P., Appourchaux, T., Boumier, P.:
{\aap} {\bf 506}, 7 
(2009)

\bibitem[\protect\citeauthoryear{Goldreich \& Keeley}{1977}]{Goldreich77}
Goldreich P., Keeley D.~A.:
\apj\ \textbf{212}, 243
(1977)

\bibitem[\protect\citeauthoryear{Gough}{1977a}]{Gough77a}
Gough D.~O.:
\apj\ \textbf{214}, 196
(1977a)

\bibitem[\protect\citeauthoryear{Gough}{1977b}]{Gough77b}
Gough D.~O.:
In Proc. IAU Colloq. No. 38, Problems of Stellar Convection,
eds  E.~A. Spiegel, J.-P. Zahn,
{\rm Lecture Notes in Physics} {\bf 71},
Springer-Verlag, Berlin, p.\,15
(1977b)

\bibitem[\protect\citeauthoryear{Gough}{1980}]{Gough80}
Gough, D. O.:
In Nonradial and Nonlinear Stellar Pulsation,
eds H.~A.\ Hill, W.~A.\ Dziembowski,
{\rm Lecture Notes in Physics} {\bf 125},
Springer-Verlag, Berlin, p.\,273
(1980)

\bibitem[\protect\citeauthoryear{Gough}{1986}]{Gough1986}
Gough, D.~O.:
In {Proc. Hydrodynamic and magnetohydrodynamic
problems in the Sun and stars}, ed.\ Y.\ Osaki, 
University of Tokyo, Tokyo, p.\,117
(1986)

\bibitem[Gough(1990)]{Gough1990} 
Gough, D.~O.:
In {Progress of Seismology of the Sun and Stars}, 
Lecture Notes in Physics {\bf 367}, eds Y.\ Osaki, H.\ Shibahashi, 
Springer Verlag, p.\,283
(1990)

\bibitem[Gough(1993)]{Gough1993b}
Gough, D.~O.:
In {Astrophysical fluid dynamics, Les Houches Session XLVII},
eds J.-P.\ Zahn, J.\ Zinn-Justin, Elsevier, Amsterdam, p.\,399 
(1993)

\bibitem[Gough(2001)]{Gough2001}
Gough D.~O.:
In {Astrophysical Ages and Timescales},
eds T.\ von Hippel, C.\ Simpson, N.\ Manset,
ASP Conf. Ser. {\bf 245},
p.\,31
(2001)

\bibitem[Gough(2002)]{Gough2002}
Gough D.~O.:
In {Stellar structure and habitable planet finding}, 
eds F.\ Favata, I.~W.\ Roxburgh, D.\ Galadi, ESA SP-485, 
ESA Publications Division, Noordwijk, p.\,65
(2002)

\bibitem[Gough \& Kosovichev(1993)]{Gough1993a}
Gough, D.~O., Kosovichev, A.~G.:
In {Proc. IAU Colloq. 137: Inside the stars},
eds A.\ Baglin, W.~W.\ Weiss,
ASP Conf. Ser. {\bf 40}, 
p.\,541 
(1993)

\bibitem[Gough {et~al.}(1996)]{Gough1996}
Gough, D.~O., Kosovichev, A.~G., Toomre, J., et al.:
{Science} {\bf 272}, 1296 
(1996)

\bibitem[Grec {et~al.}(1983)]{Grec1983}
Grec, G., Fossat, E., Pomerantz, M.:
{\solphys} {\bf 82}, 55 
(1983)

\bibitem[Grevesse \& Noels(1993)]{Greves1993}
Grevesse, N., Noels, A.:
In {Origin and evolution of the Elements},
eds N.\ Prantzos, E.\ Vangioni-Flam, M.\ Cass\'e 
Cambridge Univ. Press, Cambridge, p.\,15 
(1993)

\bibitem[Grevesse {et~al.}(2010)]{Greves2010}
Grevesse, N., Asplund, M., Sauval, J., Scott, P., Klein, O.:
{\apss} 
(2010, this volume)

\bibitem[Grundahl {et~al.}(2009)]{Grunda2009}
Grundahl, F., Christensen-Dalsgaard, J., Kjeldsen, H., J{\o}rgensen, U.~G.,
Arentoft, T., Frandsen, S., Kj{\ae}rgaard, P.:
In {Proc. GONG2008/SOHO21 meeting:
Solar-stellar Dynamos as revealed by Helio- and Asteroseismology},
eds M.\ Dikpati, T.\ Arentoft, I.\ Gonz\'alez Hern\'andez, C.\ Lindsey, 
F.\ Hill,
ASP Conf. Ser.,
{\tt [arXiv:0908.0436v1 [astro-ph.SR]]}
(2009, in the press)

\bibitem[Guzik(2006)]{Guzik2006}
Guzik, J.~A.:
In {Proc. SOHO 18 / GONG 2006 / HELAS I Conf.
Beyond the spherical Sun},
ed.\ K.\ Fletcher, ESA SP-624, ESA Publications Division,
Noordwijk, The Netherlands
(2006)

\bibitem[Guzik(2008)]{Guzik2008}
Guzik, J.~A.:
{\memsai} {\bf 79}, 481 
(2008)

\bibitem[Harvey(1988)]{Harvey1988}
Harvey, J.~W.:
In {Proc. IAU Symposium No 123, Advances in helio- and asteroseismology},
eds J.\ Christensen-Dalsgaard, S.\ Frandsen,
Reidel, Dordrecht, p.\,497 
(1988)

\bibitem[Houdek(1996)]{Houdek96}
Houdek G.:
Ph.D.\ Thesis, Pulsation of Solar-Type stars, 
University of Vienna, Vienna
(1996)

\bibitem[Houdek(2006)]{Houdek2006}
Houdek, G.:
In {Proc. SOHO 18 / GONG 2006 / HELAS I Conf.
Beyond the spherical Sun},
ed.\ K.\ Fletcher, ESA SP-624, ESA Publications Division,
Noordwijk, The Netherlands
(2006)

\bibitem[Houdek \& Gough(2007)]{Houdek2007}
Houdek G., Gough D.~O.:
\mnras\ {\bf 375}, 861
(2007)

\bibitem[Houdek \& Gough(2008)]{Houdek2008}
Houdek G., Gough D.~O.:
In {Proc.\ IAU Symp.\ 252, The Art of Modelling
Stars in the 21st Century}, eds L.\ Deng, K.~L.\ Chan, C.\ Chiosi,
Cambridge Univ. Press, Cambridge, p.\,149
(2008)

\bibitem[Houdek {et~al.}(1999)]{Houdek1999}
Houdek, G., Balmforth, N.~J., Christensen-Dalsgaard, J., Gough, D.~O.:
{\aap} {\bf 351}, 582 
(1999)

\bibitem[Kjeldsen {et~al.}(1995)]{Kjelds1995}
Kjeldsen, H., Bedding, T.~R., Viskum, M., Frandsen, S.:
{\aj} {\bf 109}, 1313 
(1995)

\bibitem[Kjeldsen {et~al.}(2003)]{Kjelds2003}
Kjeldsen, H., Bedding, T.~R., Baldry, I.~K., et al.:
{\aj} {\bf 126}, 1483 
(2003)

\bibitem[Kjeldsen {et~al.}(2005)]{Kjelds2005}
Kjeldsen, H., Bedding, T.~R., Butler, R.~P., Christensen-Dalsgaard, J.,
Kiss, L.~L., McCarthy, C., Marcy, G.~W., Tinney, C.~G., Wright, J.~T.:
{\apj} {\bf 635}, 1281 
(2005)

\bibitem[Kjeldsen {et~al.}(2008)]{Kjelds2008}
Kjeldsen, H., Bedding, T.~R., Christensen-Dalsgaard, J.:
{\apj} {\bf 683}, L175 
(2008)

\bibitem[Kjeldsen {et~al.}(2009)]{Kjelds2009}
Kjeldsen, H., Bedding, T.~R., Christensen-Dalsgaard, J.:
In {Proc. IAU Symp. 253, Transiting Planets},
eds F.\ Pont, D.\ Sasselov, M.\ Holman,
IAU and Cambridge Univ. Press, 309 
(2009)

\bibitem[Lebreton {et~al.}(2008a)]{Lebret2008a}
Lebreton, Y., Monteiro, M.~J.~P.~F.~G., Montalb\'an, J., et al.:
{\apss} {\bf 316}, 1 
(2008a)

\bibitem[Lebreton {et~al.}(2008b)]{Lebret2008b}
Lebreton, Y., Montalb\'an, J., Christensen-Dalsgaard, J., Roxburgh, I.~W.,
Weiss, A.:
{\apss} {\bf 316}, 187 
(2008b)

\bibitem[Li {et~al.}(2002)]{Li2002}
Li, L.~H., Robinson, F.~J., Demarque, P., Sofia, S.:
{\apj} {\bf 567}, 1192 
(2002)

\bibitem[Mao {et~al.}(2009)]{Mao2009}
Mao, D., Mussack, K., D\"appen, W.:
{\apj} {\bf 701}, 1204 
(2009)

\bibitem[Matthews(2007)]{Matthe2007}
Matthews, J. M.:
{\rm Comm. in Asteroseismology} {\bf 150}, 333 
(2007)

\bibitem[Matthews {et~al.}(2004)]{Matthe2004}
Matthews, J.~M., Kuschnig, R., Guenther, D.~B., Walker, G.~A.~H.,
Moffat, A.~F.~J., Rucinski, S.~M., Sasselov, D., Weiss, W.~W.:
{\nat} {\bf 430}, 51 
(Erratum: {\nat} {\bf 430}, 921)
(2004)

\bibitem[Michel {et~al.}(2008a)]{Michel2008a}
Michel, E., Baglin, A., Weiss, W.~W., et al.:
{Comm. in Asteroseismology} {\bf 156}, 73 
(2008a)

\bibitem[Michel {et~al.}(2008b)]{Michel2008b}
Michel, E., Baglin, A., Auvergne, M., et al.:
{Science} {\bf 322}, 558 
(2008b)

\bibitem[Miglio \& Montalb\'an(2005)]{Miglio2005}
Miglio, A., Montalb\'an, J.:
{\aap} {\bf 441}, 615 
(2005)

\bibitem[Monteiro \& Thompson(1998)]{Monteiro1998}
Monteiro M.~J.~P.~F.~G., Thompson M.\ J.:
In {Proc. IAU Symp.\ 185, New Eyes to see inside the Sun and Stars},
eds F.-L.\ Deubner, J.\ Christensen-Dalsgaard, D.~W.\ Kurtz, 
Kluwer, Dordrecht, p.\,317
(1998)

\bibitem[Monteiro \& Thompson(2005)]{Monteiro2005}
Monteiro M.~J.~P.~F.~G., Thompson M.~J.:
\mnras\ 361, 1187
(2005)

\bibitem[Mosser {et~al.}(2008)]{Mosser2008}
Mosser, B., Appourchaux, T., Catala, C., Buey, J.-T.\ and the 
SIAMOIS team:
In {Proc. HELAS II International Conference: Helioseismology,
Asteroseismology and the MHD Connections},
eds L.\ Gizon, M.\ Roth,
{J.\ Phys.: Conf. Ser.} {\bf 118}, 012042 
(2008)

\bibitem[Moya {et~al.}(2008)]{Moya2008}
Moya, A., Christensen-Dalsgaard, J., Charpinet, S., et al.:
{\apss} {\bf 316}, 231 
(2008)

\bibitem[Mussack \& D\"appen(2010)]{Mussac2010}
Mussack, K., D\"appen, W.:
{\apss} 
(2010, this volume)

\bibitem[Osaki(1975)]{Osaki1975}
Osaki, Y.:
{\pasj} {\bf 27}, 237 
(1975)

\bibitem[Ot\'{\i} Floranes {et~al.}(2005)]{Oti2005}
Ot\'{\i} Floranes, H., Christensen-Dalsgaard, J., Thompson, M.~J.:
{\mnras} {\bf 356}, 671 
(2005)

\bibitem[Popielski \& Dziembowski(2005)]{Popiel2005}
Popielski, B.~L., Dziembowski, W.~A.:
{\actaa} {\bf 55}, 177 
(2005)

\bibitem[Rosenthal {et~al.}(1995)]{Rosenthal95}
Rosenthal, C.~S., Christensen-Dalsgaard, J., Houdek, G.,
Monteiro, M.~J.~P.~F.~G., Nordlund, \AA., Trampedach, R.:
In Proc. 4th SOHO Workshop: Helioseismology,
eds J.~T.\ Hoeksema, V.\ Domingo, B.\ Fleck, B.\ Battrick,
ESA {SP-376}, vol.2, ESTEC, Noordwijk, p.\,459
(1995)

\bibitem[Rosenthal {et~al.}(1999)]{Rosent1999}
Rosenthal, C.~S., Christensen-Dalsgaard, J., Nordlund, {\AA}.,
Stein, R.~F., Trampedach, R.:
{\aap} {\bf 351}, 689 
(1999)

\bibitem[Roxburgh(2004)]{Roxbur2004}
Roxburgh, I.~W.:
In {Proc. 2nd Eddington workshop, ``Stellar structure and habitable
planet finding''}, ESA SP-538,
eds F.\ Favata, S.\ Aigrain,
ESA Publications Division, Noordwijk, The Netherlands, p.\,23 
(2004)

\bibitem[Roxburgh(2005)]{Roxbur2005}
Roxburgh, I.~W.:
{\aap} {\bf 434}, 665 
(2005)

\bibitem[Roxburgh(2009)]{Roxbur2009}
Roxburgh, I.~W.:
{\aap} {\bf 493}, 185 
(2009)

\bibitem[Roxburgh \& Vorontsov(1994)]{Roxbur1994}
Roxburgh, I.~W., Vorontsov, S.~V.:
{\mnras} {\bf 267}, 297 
(1994)

\bibitem[Roxburgh \& Vorontsov(2000)]{Roxbur2000}
Roxburgh, I.~W., Vorontsov, S.~V.:
{\mnras} {\bf 317}, 141 
(2000)

\bibitem[Roxburgh \& Vorontsov(2003)]{Roxbur2003}
Roxburgh, I.~W., Vorontsov, S.~V.:
{\aap} {\bf 411}, 215 
(2003)

\bibitem[Roxburgh \& Vorontsov(2007)]{Roxbur2007}
Roxburgh, I.~W., Vorontsov, S.~V.:
{\mnras} {\bf 379}, 801 
(2007)

\bibitem[Salpeter(1954)]{Salpet1954}
Salpeter, E.~E.:
{Austr. J.\ Phys.} {\bf 7}, 373 
(1954)

\bibitem[Shaviv(2004)]{Shaviv2004}
Shaviv, G.:
In {Equation-of-State and Phase-Transition Issues in
Models of Ordinary Astrophysical Matter},
eds V.\ {\v C}elebonovi\'c,  W.\ D\"appen, D.\ Gough,
AIP Conf. Proc. {\bf 731}, AIP, Melville, New York, p.\,67 
(2004)

\bibitem[Stello {et~al.}(2009)]{Stello2009}
Stello, D., Chaplin, W.~J., Bruntt, H., et al.:
{\apj} {\bf 700}, 1589 
(2009)

\bibitem[Straka {et~al.}(2006)]{Straka2006}
Straka, C.~W., Demarque, P., Guenther, D.~B., Li, L., Robinson, F.~J.:
{\apj} {\bf 636}, 1078 
(2006)

\bibitem[Tassoul(1980)]{Tassoul1980}Tassoul M.:
{\apjs} {\bf 43}, 46
(1980)

\bibitem[Ulrich(1986)]{Ulrich1986}
Ulrich, R.~K.:
{\apj} {\bf 306}, L37 
(1986)

\bibitem[Vandakurov(1967)]{Vandakurov1967}
Vandakurov, Y.~V.:
{\azh} {\bf 44}, 786
(English translation: {\sovast} {\bf 11}, 630)
(1967)

\bibitem[VandenBerg {et~al.}(2007)]{Vanden2007}
VandenBerg, D.~A., Gustafsson, B., Edvardsson, B., Eriksson, K.,
Ferguson, J.:
{\apj} {\bf 666}, L105 
(2007)

\bibitem[Walker {et~al.}(2003)]{Walker2003}
Walker, G., Matthews, J., Kuschnig, R., et al.:
{\pasp} {\bf 115}, 1023 
(2003)

\bibitem[Weiss {et~al.}(2001)]{Weiss2001}
Weiss, A., Flaskamp, M., Tsytovich, V.~N.:
{\aap} {\bf 371}, 1123 
(2001)


\end{thebibliography}

\end{document}